\documentclass[conference]{IEEEtran}
\pdfoutput=1
\IEEEoverridecommandlockouts
\usepackage{lipsum}
\usepackage{cite}
\usepackage{amsmath,amssymb,amsfonts}
\usepackage{algorithmic}
\usepackage{graphicx}
\usepackage{hyperref}
\usepackage{bm}
\hypersetup{
colorlinks=true,
linkcolor=black,
urlcolor=black,
citecolor=black}
\usepackage{subfigure}
\usepackage{blindtext}
\usepackage{balance}
\usepackage{textcomp}
\usepackage{xcolor}
\usepackage{makecell}
\usepackage{multirow}
\usepackage[marginal]{footmisc} 
\renewcommand{\footnoterule}{
  \kern -3pt
  \hrule width 2in
  \kern 2pt
}
\newcommand\blfootnote[1]{%
\begingroup
\renewcommand\thefootnote{}\footnote{#1}%
\addtocounter{footnote}{-1}%
\endgroup
}

\usepackage[export]{adjustbox} 
\def\BibTeX{{\rm B\kern-.05em{\sc i\kern-.025em b}\kern-.08em
    T\kern-.1667em\lower.7ex\hbox{E}\kern-.125emX}}

\title{
OLxPBench: Real-time, Semantically Consistent, and Domain-specific are Essential in Benchmarking, Designing, and Implementing HTAP Systems
}

\author{\IEEEauthorblockN{Guoxin Kang$^{1, 2}$,
Lei Wang$^{1,3}$, Wanling Gao$^{1,3}$, 
Fei Tang$^{1, 2}$, and Jianfeng Zhan$^{1, 2,3}$}

\IEEEauthorblockA{$^{1}$ Research Center for Advanced Computer Systems,\\
Institute of Computing Technology, Chinese Academy of Sciences\\ \{gaowanling, wanglei\_2011, zhanjianfeng\}@ict.ac.cn}
\IEEEauthorblockA{$^{2}$ University of Chinese Academy of Sciences\\ \{kangguoxin, tangfei\}@ict.ac.cn}
\IEEEauthorblockA{$^{3}$ International Open Benchmark Council}}

\begin{document}

\maketitle

\begin{abstract}
As real-time analysis on the fresh data become increasingly compelling, more organizations deploy Hybrid Transactional/Analytical Processing (HTAP) systems to support real-time queries on data recently generated by online transaction processing.  This paper argues that real-time queries, semantically consistent schema, and domain-specific workloads are essential in benchmarking, designing, and implementing HTAP systems.   
However, most state-of-the-art and state-of-the-practice benchmarks ignore those critical factors. Hence, at best, they are incommensurable and, at worst, misleading in benchmarking, designing, and implementing HTAP systems.
This paper presents OLxPBench, a composite HTAP benchmark suite.
OLxPBench proposes: (1) the abstraction of a hybrid transaction, performing a real-time query in-between an online transaction, to model widely-observed behavior pattern -- making a quick decision while consulting real-time analysis; (2) a semantically consistent schema to express the relationships between OLTP and OLAP schema;  (3) the combination of domain-specific and general benchmarks to characterize diverse application scenarios with varying resource demands.
Our evaluations justify the three design decisions of OLxPBench and pinpoint the bottlenecks of two mainstream distributed HTAP DBMSs.
International Open Benchmark Council (BenchCouncil) sets up the OLxPBench homepage at ~\url{https://www.benchcouncil.org/olxpbench/}. Its source code is available from ~\url{https://github.com/BenchCouncil/olxpbench.git}.
\blfootnote{$\ddagger$\textbf{Jianfeng~Zhan~is~the~corresponding~author.}}
\end{abstract}


\begin{figure}[!t]
 \centering
 \includegraphics[scale=0.38]{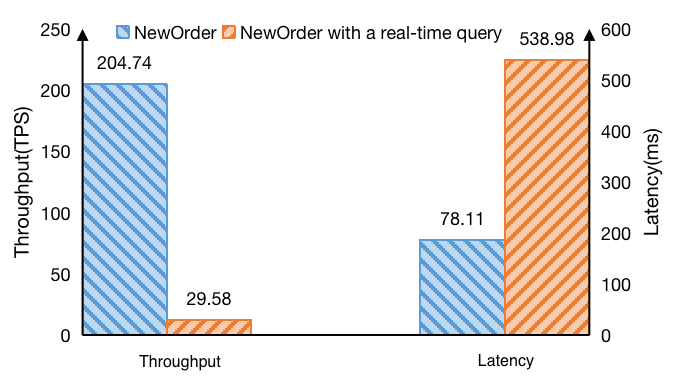}
 \caption{This figure reveals the impact of a hybrid workload -- performing a real-time query in-between an online transaction on the performance of TiDB -- a state-of-the-art HTAP system against that of only an online transaction. The significant performance gap justifies why we should consider real-time queries in benchmarking HTAP systems. The other two critical factors that the HTAP benchmarks must consider are semantically consistent schema and domain-specific workloads.}
 \label{fig: 1}
 \vspace{-0.6cm}
\end{figure}

\section{Introduction}
In recent years, hybrid transaction/analytical processing (in short, HTAP) systems are proliferating fast.
Many giant companies provide HTAP systems, including Oracle database in-memory~\cite{lahiri2015oracle}, SQL Server~\cite{larson2015real}, SAP HANA~\cite{lee2017parallel}, MemSQL~\cite{MemSQL111} and TiDB~\cite{huang2020tidb}.
HTAP systems are popular for two reasons.
First, giant companies are demanding fresher data in shorter shipping duration for real-time customer analysis~\cite{harvard111}. 
Throughout this paper, real-time emphasizes performing a task like data analysis or user behavior simulation interactively in contexts like financial fraud monitoring or  recommendation~\cite{pi2020search}~\cite{ma2020temporal}. Our usage of real-time is different from its traditional definition that limits the operations to completion within a hard or soft deadline~\cite{realtime111}.
The value of mass business data will diminish with time~\cite{pavlo2016s}.
Moving data from the OLTP system to the OLAP system is complex and time-consuming.
Meanwhile, it is impossible to perform real-time analysis on data that has passed a long turnaround time.
Besides, software development and maintenance for two separate systems are also expensive.
Second, the advanced modern software and hardware technologies, including in-memory computing (IMC) techniques~\cite{IMC}, multi-core processors, various levels of memory caches, and large memories~\cite{bog2012interactive}~\cite{ozcan2017hybrid}, contribute to the HTAP systems' rapid development.

There are three primary architectures for HTAP systems design and implementation. 
The first one introduces extract-transform-load (ETL) processing~\cite{vassiliadis2002conceptual} between the OLTP DBMS and the data warehouse to complete data migration, data format transformation, and complex data analysis. However, the ETL systems are not competent for real-time analysis as they introduce time and space costs that can not be neglected. The second one is to utilize a stream processing system~\cite{toshniwal2014storm}~\cite{zaharia2013discretized} that feeds the incoming data to a batch processing system ~\cite{shvachko2010hadoop}~\cite{armbrust2015spark}.
Due to several unfavorable factors such as strong consistency semantics and double operating cost, it is not easy to use separate stream processing systems as HTAP solutions.
The other architecture uses a single HTAP DBMS ~\cite{kemper2011hyper, lee2017parallel, makreshanski2017batchdb, huang2020tidb}, achieving high performance for OLTP and OLAP.
It eliminates data movement overhead but adds pressure to performance isolation or consistency guarantees. In the face of numerous HTAP solutions, it is hard to compare different system performances.
We describe these three architectures in detail in Section~\ref{sec: 3.1}.

The HTAP benchmarks provide quantitative metrics, methodology, and tools to evaluate different systems and their specific design decisions, offering valuable design and implementation inputs. However, as summarized in Table~\ref{table: 4},  most state-of-the-art~\cite{coelho2017htapbench} and state-of-the-practice~\cite{cole2011mixed} HTAP benchmarks fail to consider real-time, semantically consistent, and domain-specific~\cite{qu2022current} in benchmarking, designing, and implementing HTAP systems. Hence, at best, they are incommensurable and, at worst, misleading in benchmarking, designing, and implementing HTAP systems. We quantitatively justify why we consider those three critical factors.

First, being real-time is essential. On the one hand, real-time is crucial to customer analysis -- the fresher the data, the higher the value. On the other hand, there are widely-observed user behavior patterns -- performing real-time analysis before making a quick decision.   For example, if the customer wants to order an item in e-commerce, a query to get the lowest price rather than the random price of the item is most likely to happen before ordering the item. We propose the abstraction of a hybrid transaction, which performs a real-time query in-between an online transaction, to model this user behavior pattern, which the previous HTAP benchmarks overlook.
Figure~\ref{fig: 1} shows the impact of a hybrid transaction on the performance of the online transactions in TiDB~\cite{huang2020tidb} -- a state-of-the-art HTAP system against that of a sole online transaction as a baseline.
The real-time query increases the baseline latency by a factor of 5.9, and decreases the baseline throughput by a factor of 5.9.

Second, the previous works use stitch schema. For example, the stitch schema in state-of-the-art HTAP benchmarks~\cite{cole2011mixed}\cite{coelho2017htapbench} just reuses the schema from TPC-C~\cite{TPCC111} and TPC-H~\cite{TPCH111}.
Instead, in real-world application scenarios, the data operated by the analytical query is generated by the online transaction, so the OLAP schema is a subset of the OLTP schema. We call this characteristic semantically consistent.  
The stitch schema can not disclose the severe interferences between analytical workload and transactional workloads in real-world scenarios. Our experiment result shows that the analytical workloads decrease transactional throughput by 89\% using the semantically consistent schema, rather than 10\% using stitch schema in the previous work~\cite{huang2020tidb}.

Last but not least, most of the previous works only provide a general HTAP benchmark. 
The generic benchmark reflects a wide range of use cases. Instead, real-world applications have different workloads and schema with varying resource demands. So we propose the domain-specific benchmarks, which are specialized to evaluate the performance of HTAP systems in one or several specific scenarios.
In Section~\ref{sec: 6}, our evaluation shows that the peak transactional throughput of a general benchmark is nearly 20 times that of a domain-specific benchmark on the same testbed.
Their performance varies greatly depending on the complexity of the relationships in the table, the read/write ratio of the transaction, and different system resource requirements.

We propose an HTAP benchmark suite in addition to a benchmarking framework named OLxPBench to help users perform performance comparisons. The main contributions of this paper are as follows.
(1) We quantitatively justify why we should consider real-time queries, semantically consistent schema, and domain-specific workloads in HTAP benchmarking~\cite{zhan2021call}. OLxPBench proposes new built-in hybrid workloads that perform a real-time query in-between an online transaction; a semantically consistent HTAP schema; one general benchmark; and two domain-specific benchmarks to evaluate the HTAP systems, including retail, banking, and telecommunications~\cite{harvard111}.

(2) We design and implement an extensible HTAP benchmarking framework and three comprehensive HTAP benchmarks: subenchmark, fibenchmark, and tabenchmark;
Compared against the most related work -- CH-benCHmark~\cite{cole2011mixed}, our work is not trivial because there are eighteen analytical queries and seventeen hybrid queries tailored for different benchmarks.
OLxPBench provides valuable experience in designing and implementing schematically consistent schema, hybrid workloads, and domain-specific benchmarks in HTAP benchmarking.

(3) Extensive experiments are conducted on the two mainstream HTAP systems: MemSQL and TiDB using OLxPBench against CH-benCHmark~\cite{cole2011mixed}. We have observed the following insights. 
The vertical table technique adopted by MemSQL is not very helpful to deal with the hybrid workloads because a large number of join operations generated by relationship query statements increases the waiting time of the hybrid transactions;
The mainstream HTAP systems have poor performance in scanning operations for composite primary keys; The mutual interference between online transactions and analytical queries causes poor performance isolation.

\section{Related Works} ~\label{sec: 2}
We compare OLxPBench with five other state-of-the-art and state-of-the-practice HTAP benchmarks in Table~\ref{table: 4}.
We classify the HTAP benchmarks into two groups according to the complexity of their workloads.
The one contains intricate transactions and queries, such as CH-benCHmark~\cite{cole2011mixed}, CBTR~\cite{bog2012interactive}, and HTAPBench~\cite{coelho2017htapbench}. 
The other includes a mix of simple insert/select operations, i.e., ADAPT~\cite{arulraj2016bridging} and HAP~\cite{athanassoulis2019optimal}.
The real-time queries generally involve simple
$aggregate$ operations and the analytical queries include more complex operations.
\begin{table*}
\centering  
\caption{Comparison of OLxPBench With State-of-the-art and State-of-the-practice Benchmarks.} 
\scalebox{0.80}{
\begin{tabular}{|c|c|c|c|c|c|c|c|c|c|} 
\hline 
Name&Online transaction&Analytical query&Hybrid transaction&Real-time query&Semantically consistent schema&General benchmark&Domain-specific benchmark\\
\hline  
CH-benCHmark&$\surd$&$\surd$&$\times$&$\times$&$\times$&$\surd$&$\times$\\
\hline 
CBTR&$\surd$&$\surd$&$\times$&$\times$&$\surd$&$\times$&$\surd$\\
\hline
HTAPBench&$\surd$&$\surd$&$\times$&$\times$&$\times$&$\surd$&$\times$\\
\hline
ADAPT&$\times$&$\times$&$\times$&$\times$&$\surd$&$\surd$&$\times$\\
\hline
HAP&$\times$&$\times$&$\times$&$\times$&$\surd$&$\surd$&$\times$\\
\hline
OLxPBench&$\surd$&$\surd$&$\surd$&$\surd$&$\surd$&$\surd$&$\surd$\\
\hline
\end{tabular}}
\label{table: 4}
\vspace{-0.5cm}
\end{table*}
 
CH-benCHmark launches the online transactions adopted from TPC-C and analytical queries from TPC-H  concurrently on the stitch schema; however, they have different business semantics. 
Moreover, CH-benCHmark never updates the Supplier, Nation, and Region tables used by OLAP since the online transactions only update partial OLAP tables using stitch schema. Thus, its OLTP and OLAP  operate different data and further cover up the contention in massive concurrency.
HTAPBench uses the same schema model with CH-benCHmark.
Besides, HTAPBench~\cite{coelho2017htapbench} implements the Client Balancer to control the number of analytical queries to avoid too many analytical queries affecting the performance of online transactions.
CBTR mimics the order-to-cash process of real enterprise systems~\cite{bog2012interactive} and provides more complex as well as dynamic workloads.
It is indisputable that CBTR has effective instruction for ad-hoc database design.
Unfortunately, CBTR does not include the real-time query and is not open-source.
Besides, its single domain-specific benchmark is insufficient to evaluate HTAP solutions in various scenarios.

ADAPT benchmark has two tables -- a narrow table and a wide table~\cite{arulraj2016bridging}. 
The operations are abstracted from a natural productive environment, and the read-only operations are 80\%. 
The HAP benchmark is based on the ADAPT benchmark and expands the update and deleted operation to test the storage engine.
The read-only operations are 50\%. Overall, the operations are too simple to represent complex transactions in natural business environments.

When designing OLxPBench, we choose the semantically consistent schema rather than the stitched schema~\cite{coelho2017htapbench}\cite{cole2011mixed} to expose the original interference between OLTP workloads and OLAP workloads.
Besides, we increase the real-time query for real-time user behavior analyzing and simulating.
We also obey both general~\cite{TPCC111} and domain-specific~\cite{seltzer1999case} principles.
The general benchmark helps designers perform performance comparisons, and the domain-specific benchmarks help the user select the HTAP DBMS that best support their specific workloads. 
Moreover, we compare the semantically consistent schema to stitched schema~\cite{cole2011mixed} in Section~\ref{sec: 5.3.1} because CH-benCHmark~\cite{cole2011mixed} originates from the HTAP transactional benchmark.

\section{Background and Motivation} ~\label{sec: 3}
\subsection{The Background of HTAP Systems} ~\label{sec: 3.1}
HTAP DBMSs need to perform trade-offs considering different performance requirements of different workloads.
Currently, there are three types of HTAP solutions,  and we compare their pros and cons in the following subsections. 

The first solution is to use separate DBMSs~\cite{raman2013db2, lahiri2013oracle, raza2020adaptive, yang2020f1} to achieve high performance of online transactions and analytical queries. Generally, online transactions adopt a row-based store due to its high efficiency for records insert and update. Analytical queries often adopt a column-based data warehouse since it supports efficient data scans. However, separate DBMS needs to convert row-based data to column-based ones, i.e., ETL processing, which is too time-consuming to analyze the latest data and make instant decisions. 
For example, Pavlo et al.~\cite{pavlo2016s}\cite{ozcan2017hybrid} refer to the standard ETL process to migrate data from the OLTP system to the OLAP system as one of the solutions to HTAP.

Second, the lambda architecture~\cite{marz2013big, kiran2015lambda, lin2017lambda}, which consists of a real-time stream processing system and a batch processing system, can perform real-time analytics on incoming data, but it is expensive.
The real-time stream processing~\cite{toshniwal2014storm} systems provide views of online data while simultaneously using batch processing~\cite{HADOOP111} to provide comprehensive and accurate views of batch data.
Besides, the serving layer merges the real-time view with the batch views and then responses to the end-user.
The lambda architecture provides a real-time analysis at a considerable cost,  including double write costs, double or more development costs, and so on.
In brief, the cost of maintaining two systems is also very high.

Third, using a single HTAP DBMS to handle online transactions and real-time queries.
Because real-time analytics on fresh data is valuable, the lightweight propagation technique is developed to transfer recent transactional logs to the analytical storage node in a more short and flexible schedule~\cite{makreshanski2017batchdb}.
Microsoft SQL Server~\cite{larson2015real} stores hot inserted and updated rows in middle delta storage to transfer them to OLAP storage and speed up query processing.
Oracle in-memory database~\cite{lahiri2015oracle} keeps a dual-format store for OLTP and OLAP workloads without double memory requirements.
It uses a read-only snapshot maintained in memory for analysis.
Their latest work~\cite{mukherjee2016fault} provides a more available distributed architecture and fault tolerance than the original.
MemSQL uses an in-memory row-store and an on-disk column-store to handle highly concurrent operational and analytical workloads~\cite{MemSQL111}. 
The storage layer of TiDB consists of a row-based store and a column-based store.
To analyze real-time queries of the fresh data, TiDB uses asynchronous log replication to keep the data consistent~\cite{huang2020tidb}.
The IBM Db2 Analytics Accelerator uses replication technology to enhance its real-time capacity~\cite{butterstein2020replication}.
VEGITO~\cite{shen2021retrofitting} retrofits the high availability mechanism to support HTAP workloads.

\subsection{Motivation} ~\label{sec: 3.2}
\subsubsection{It is mandatory to include real-time queries in HTAP benchmarking\label{sec: 3.2.1}}

HTAP benchmarking should contain real-time queries for the following reasons.
First, real-time queries matter in customer analysis.
HTAP DBMSs enable more informed and in-business real-time decision-making~\cite{MGHIC}.
Real-time customer analysis is crucial because it is the basis of instant decision-making and fraud detection.
The fresher the data, the higher the value.
Real-time queries are usually executed on the recent data committed by transactions.
For example, if an item requested by a customer has been sold out according to the real-time inventory, similar ones will be recommended instantly.

Second, real-time queries can be used to mimic real-time user behavior.
For example, if a customer wants to create a ${New\_Order}$ transaction in TPC-C~\cite{TPCC111}, what is most likely to happen before selecting an item during the ${New\_Order}$ transaction~\cite{cole2011mixed} is a real-time query that finds the lowest price of the goods, rather than the random price.
However, none state-of-the-art~\cite{coelho2017htapbench} and state-of-the-practice~\cite{cole2011mixed} HTAP benchmarks provide the workloads that include real-time queries imitating user behavior.

Figure~\ref{fig: 1} shows the impact of a real-time query on the performance of TiDB~\cite{huang2020tidb} -- a state-of-the-art HTAP system. The ${New\_Order}$ transaction is the same as the ${New\_Order}$ transaction in TPC-C~\cite{TPCC111}. The real-time query is an aggregate operation that gets the item's lowest price in real time.
Real-time queries in the subenchmark are from a top-tier E-commerce internet service provider.
The experimental setup is the same as Section~\ref{sec: 5.1}.
When a real-time query is injected in the ${New\_Order}$ transaction~\cite{coelho2017htapbench}~\cite{cole2011mixed}, the average latency increases by 5.9x, and the throughput reduces by 5.9x.
So it is mandatory to include real-time queries in the HTAP benchmarks, or else the evaluation result will be misleading~\cite{zhan2021call}.

\subsubsection{Semantically consistent schema is essential\label{sec: 3.2.2}}
\begin{table*}[t]
\centering  
\caption{Features of the OLxPBench workloads.} 
\scalebox{0.89}{
\begin{tabular}{c c c c c c c c c}
\hline 
Benchmark&Tables&Columns&Indexes&OLTP Transactions&Read-only OLTP Transactions&Queries&Hybrid Transactions&Read-only Hybrid  Transactions\\
\hline  
Subenchmark&9&92&3&5&8.0\%&9&5&60.0\%\\
Fibenchmark&3&6&4&6&15.0\%&4&6&20.0\%\\
Tabenchmark&4&51&5&7&80.0\%&5&6&40.0\%\\
\hline
\end{tabular}}
\label{table: 3}
\vspace{-0.5cm}
\end{table*}
The state-of-the-art~\cite{coelho2017htapbench} and state-of-the-practice~\cite{cole2011mixed} HTAP benchmarks all use stitch schema.
The stitch schema integrates the TPC-C schema~\cite{TPCC111} with TPC-H~\cite{TPCH111} schema and includes 12 tables.
The ${NEW-ORDER}$, ${STOCK}$, ${CUSTOMER}$, ${ORDERLINE}$, ${ORDERS}$, and ${ITEM}$ tables are accessed by TPC-C and TPC-H.
The ${WAREHOUSE}$, ${DISTRICT}$, and ${HISTORY}$ tables are  only accessed by TPC-C. The ${SUPPLIER}$, ${NATION}$, and ${REGION}$ tables are accessed by TPC-H only.
Both TPC-C and TPC-H keep the third normal form to reduce data duplication. 

There are two flaws in such a stitch schema: 
First, OLTP and OLAP operate on the same business data in real scenarios; however, the query only analyzes one-sided business data with the stitch schema, leading to biased decisions.
For example, in CH-benCHmark, the stitch schema only allows queries to analyze data from the shared six tables between TPC-C and TPC-H.
When the ${Payment}$ transaction in CH-benCHmark is completed, a record will be written in the history table.
The records in the history table are essential for analyzing the custom's behavior.
However, none of the analysis queries in previous benchmarks~\cite{coelho2017htapbench}~\cite{cole2011mixed}  can analyze the tens of thousands of records in the history table.
In addition, there is no query to analyze the warehouse table and district table of TPC-C~\cite{coelho2017htapbench} ~\cite{cole2011mixed}.
It is very costly to discard valuable parts of OLTP data.
The stitch schema leads the results of the analytical queries to be partial, perplexing, and incorrect.

Second, with stitch schema, the competitions between analytical workload and transactional workloads are hidden, making it impossible to fairly evaluate the interference between analytical workload and transactional workloads in real-world scenarios.
The online transactions and analytical queries operate on the same business data in the real world, so intense competitions for resources are not avoidable. 
However, in the previous benchmark~\cite{coelho2017htapbench, cole2011mixed}, 45.4\%, 40.9\%, and 13.6\% of the 22 queries on the stitch schema access the ${SUPPLIER}$, ${NATION}$, and ${REGION}$ tables that never update or insert records, respectively.
The low competition between analytical and transactional workload will propagate a false image that the HTAP system can guarantee the isolated performance for separate OLTP and OLAP workloads~\cite{huang2020tidb}.
In Section~\ref{sec: 6.1.2}, we use the general benchmark in OLxPBench, which uncovers the high competition between the OLTP and OLAP workloads, to evaluate the TiDB and find the throughput interference of OLTP and OLAP is as high as 89\% and 59\%, respectively.

\subsubsection{Domain-specific benchmarks should be included\label{sec: 3.2.3}}
CH-benCHmark~\cite{cole2011mixed} is a general benchmark that fails to evaluate the HTAP system performance in a particular application scenario.
In Section~\ref{sec: 6}, our evaluation shows that the peak transactional throughput of a general benchmark is nearly 20 times that of a domain-specific benchmark on the same testbed.
In the face of numerous HTAP solutions, there is an urgent need to consider generic and domain-specific HTAP DBMS benchmarks.
OLxPBench provides one generic benchmark and two domain-specific benchmarks for evaluating HTAP systems.
In Sections~\ref{sec: 6.2.1} and~\ref{sec: 6.3.2}, our evaluation shows that the peak transactional throughput of a domain-specific benchmark is nearly 200 times that of another domain-specific benchmark on the same testbed.
OLxPBench implements an extensible framework that makes it easy for developers to add a new benchmark.

\section{The design and implementation~\label{sec: 4}}
To fully evaluate HTAP DBMSs, we present OLxPBench, consisting of a general benchmark and two domain-specific benchmarks. 
The general benchmark, which we name subenchmark, extracts complex operations from the retail activity and does not attempt to model an actual application scenario~\cite{TPCC111}, which intends to perform performance comparison for HTAP DBMSs. 
Meanwhile,  OLxPBench has two domain-specific benchmarks, which we name fibenchmark and tabenchmark, model the financial~\cite{alomari2008cost} and telecommunication~\cite{TATP111} scenarios, and help the users select the HTAP DBMS that best support their specific workloads.
This section introduces the OLxPBench suite from the schema model design, the details of workloads, the benchmark categories, and implementation.

\subsection{HTAP Schema model design}~\label{sec: 4.1}
We follow three principles in designing the HTAP schema.

(1) Any record accessible to OLTP should be accessible to OLAP.
Because online transactions generate the data that the analytical queries will analyze.
The OLTP schema set should include the OLAP schema.
We first propose that in the HTAP benchmarks,  the mixed workloads, including OLTP and OLAP workloads, should use the semantically consistent schema. They will reveal the inherent interference between OLTP workload and OLAP workloads.

(2) The schema models should be diverse and practical for thoroughly evaluating the various HTAP solutions.
The diversity of schema models is reflected in the diversity of practical uses.
Therefore, we provide the generic schema model for performance comparisons and two domain-specific schema models for users to select the HTAP DBMS that best support their specific workloads.
We choose schema models for retail, banking, and telecommunications activities because the above practitioners were among the first to adopt the HTAP solutions~\cite{harvard111}.

(3) The design of integrity constraints should be relevant to implementing a specific HTAP database.  
For example, some HTAP DBMSs (such as MemSQL~\cite{MemSQL111}) do not currently support foreign keys.
As a result, OLxPBench's schema models come in two versions, one with no fundamental foreign constraint and one with foreign constraint, which users can choose on-demand.

\subsection{The details of HTAP Workloads}~\label{sec: 4.2}

OLxPBench contains nine built-in workloads with different types and complexity.
Three online transaction workloads are extracted from popular benchmarks~\cite{TATP111}~\cite{TPCC111}~\cite{cahill2009serializable}.
In addition, we add three analytical query workloads and three hybrid transaction workloads for real-time customer analysis and simulating the real-time user behaviors.
We distill the E-commerce services from an industry partner, which we keep anonymously at its request, into representative real-time queries. 
The analytical workloads contain complex analytical operations such as $multi-join$, $sub-selection$, $Group-By$, and $Order-By$ operations based on the different schema models.
Table~\ref{table: 3} describes the features of these benchmarks.

In more specific implementations, we modify the integrity constraints of the schema of SmallBank~\cite{alomari2008cost} and TATP~\cite{TATP111} to adapt the implementation of the MemSQL~\cite{MemSQL111}.
Furthermore, we increase the composite primary key to TATP~\cite{TATP111}, which is common in real business scenarios.
OLxPBench provides valuable experience for schema model design and hybrid transaction abstraction.
The request rates, transaction/query weights, and schema relations are configurable for different testing purposes.
This subsection will introduce the details of these benchmarks in turn.

\subsubsection{Subenchmark} ~\label{sec: 4.2.1}
The subenchmark is inspired by TPC-C~\cite{TPCC111}, which is not bound to a specific scenario, and the community considers a general benchmark for OLTP system evaluation.
The online workloads of the subenchmark are the same as TPC-C's transactions, which are write-heavy and merely 8\% read-only transactions.
The online transactions include ${NewOrder}$, ${Payment}$, ${OrderStatus}$, ${Delivery}$, and ${StockLevel}$.

The nine analytical queries in the subenchmark keep the essential characteristics such as the complexity of operations. The analytical queries perform multi-join, aggregation, grouping, and sorting operations on a semantically consistent schema.
For example, the Orders Analytical Report Query (Q1) is designed for getting the magnitude summary for all $\textit{ORDER\_LINE}$ items as of a given date.
The query lists the total quantity, total amount, average quantity, and average amount for further analysis.
The above aggregates are grouped by their number and listed in ascending order.
We newly increase five hybrid transactions, and the default configuration of the subenchmark has 60\% read-only hybrid transactions.
The real-time queries that simulate the user behavior are the representative aggregation operations in the actual E-commerce production application: if the customer wants to create a ${New\_Order}$ transaction, a query to get the lowest price rather than the random price of the item (X1).

\subsubsection{Fibenchmark}~\label{sec: 4.2.2}
The fibenchmark is inspired by SmallBank~\cite{alomari2008cost},  which aims at bank scenarios.
Hence, it is a domain-specific benchmark.
The fibenchmark contains three tables: $\textit{ACCOUNT}$, $\textit{SAVING}$, and $\textit{CHECKING}$, and the transactions mainly modify the customers' accounts.
The online transactions are $\textit{Amalgamate}$, $\textit{Balance}$, $\textit{DepositChecking}$, $\textit{SendPayment}$, $\textit{TransactSavings}$, and $\textit{WriteCheck}$.
Fifteen percent of the above transactions are real-only in the default configuration.
We keep all the online transactions of SmallBank~\cite{alomari2008cost}, and we newly increase the analytical workloads and the hybrid transactions in fibenchmark.
 
The analytical workloads perform real-time customer account analytics.
The complex queries include ${join}$, ${aggregate}$, ${sub-selection}$, ${Order-By}$ and ${Group-By}$ operations.
For example, the Account Name Query (Q1) lists the name in the combining row from $\textit{ACCOUNT}$ and $\textit{CHECKING}$ tables.
Besides, the real-time queries in hybrid transactions are generally the ${aggregate}$ operations and perform the real-time financial analysis on the user's account.
There are six hybrid transactions, and the default configuration of the fibenchmark has 20\% read-only hybrid transactions.
For example, the Checking Balance Transactions (X6) checks whether the cheque balance is sufficient and aggregates the value of the minimum savings.
The volatility of extreme values is also an important research topic in the financial field.

\begin{figure}[!t]
 \centering
 \includegraphics[scale=0.30]{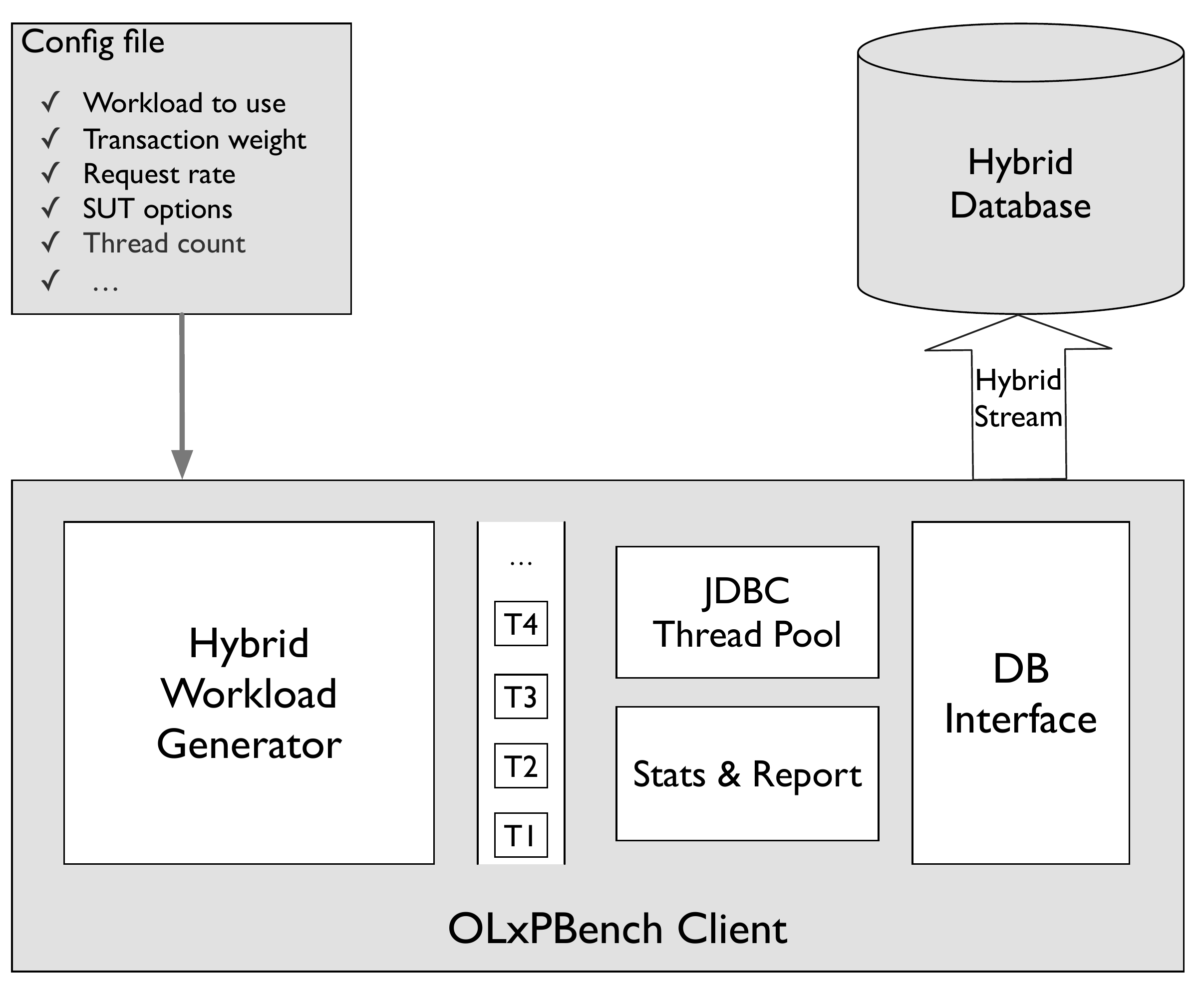}
 \caption{OLxPBench architecture.}
 \label{fig: 2}
 \vspace{-0.5cm}
\end{figure}

\subsubsection{Tabenchmark} ~\label{sec: 4.2.3}
The tabenchmark is inspired by TATP~\cite{TATP111}, which aims at telecom scenarios.
Hence, it is a domain-specific benchmark.
The online transactions in the tabenchmark simulate a typical Home Location Register (HLR) database used by a mobile carrier~\cite{TATP111}.
Eighty percent of online transactions are real-only and the transactions are $\textit{DeleteCallForwarding}$, $\textit{GetAccessData}$, $\textit{GetNewDestination}$, $\textit{GetSubscriberData}$, $\textit{InsertCallForwarding}$, $\textit{UpdateLocation}$ and $\textit{UpdateLocation}$.
We modify the primary key of the ${SUBSCRIBER}$ table from ${s\_id}$ to ${(s\_id, sf\_type)}$, because the composite primary key is standard in the real business scenario.
The original data definition language file is a choice.
We keep all the online transactions of TATP~\cite{TATP111}, and we newly increase the analytical workloads and the hybrid transactions in fibenchmark.

The analytical queries help the mobile network operators to analyze user behavior in real-time.
The analytical queries also comprise the ${arithmetic}$ operation besides the operations in the fibenchmark.
For example, the Start Time Query (Q3) calculates the average of the starting time of the call forwarding. 
The average value of start time is essential for load forecasting.
The real-time queries perform the real-time activities on practical mobile users.
The real-time query not only performs aggregation operation but also does a fuzzy search based on the sub-string.
For example, the Fuzzy Search Transaction (X6) queries all information about the subscriber. 
It selects the subscriber IDS whose user data matches the fuzzy search criteria.

\subsection{The implementation of OLxPBench}~\label{sec: 4.3}

OLxPBench is used for evaluating distributed HTAP DBMSs and other HTAP DBaaS systems that support SQL through JDBC. The architecture of the OLxPBench is shown in Figure~\ref{fig: 2}. OLxPBench parses configuration files at runtime and generates the corresponding hybrid workloads. Then the hybrid workloads are populated in request queues.
The request rates, transaction types, real-time query types, weights, and target DB configuration are specified in the XML file.
The thread connects to the target database by JDBC pool and pulls requests from the request queue.
The threads also measure the latency and throughput metrics.
Finally, the statistics module aggregates the above metrics and stores the min, max, medium, 90th, 95th, 99.9th, and 99.99th percentile latency in a file specified by the user in the terminal.

The open-loop mode sends the requests with the precise request rate control mechanism because the open-loop load generator sends the request without waiting for the previous request to come back. 
However, in a closed-loop mode, the response of a request triggers the sending of a new request.
Besides, the users can customize the weights of various online transactions and analytical queries.
OLxPBench is inspired by the OLTP-Bench's OLTP module~\cite{difallah2013oltp} and newly increased analytic and hybrid modules.
OLxPBench achieves three online and analytical agent combination modes for different HTAP solutions.
The first mode sequentially sends the online transactions or analytical queries.
The second mode concurrently invokes the transactional workload and analytical workload.
The last mode sends hybrid transactions performing a real-time query in-between an online transaction to simulate the user behavior.
The OLxPBench client is a java program and is easy to extend with new hybrid database back-ends.

\section{Evaluation}~\label{sec: 5}
Our evaluation illustrates the effectiveness of OLxPBench. In Section~\ref{sec: 5.3}, we compare OLxPBench with the state-of-the-practice~\cite{cole2011mixed} work, testing the key features of OLxPBench and reporting the standard deviation of absolute value. From Section~\ref{sec: 6.1} to Section~\ref{sec:find}, we evaluate the mainstream distributed HTAP DBMSs using OLxPBench and pinpoint the bottlenecks of two mainstream distributed HTAP DBMSs. In Section~\ref{sec: 6.3}, we evaluate the scaling capability of TiDB, MemSQL, and OceanBase.

\subsection{Experimental Setup~\label{sec: 5.1}}
\subsubsection{Cluster Deployment}~\label{sec: 5.1.1}
We deploy a 4-node cluster for our evaluation. Each server includes 2 Intel Xeon E5-2620@2.40GHz CPUs, 64 GB memory, and 1TB SSD.
Each CPU has 6 physical cores, and hyper-thread is enabled.
We used all of the 24 hardware threads. 
For the scaling capability experiments in Section~\ref{sec: 6.3}, we deploy a 16-node cluster, with each cloud server including 8 Intel Xeon Platinum 8269CY@2.50GHz virtual CPUs, 32 GiB memory, and 140GiB enhanced solid-state disk (ESSD).
We used all of the 8 threads.
All machines are configured with Intel Ethernet 1GbE NICs.
The operating system is ubuntu 16.04 with the Linux kernel 4.4.

\subsubsection{Database Deployment} ~\label{sec: 5.1.1}
In our experiments, 4-node configuration ensures that the components of systems are under test are distributed deployed.
TiDB~\cite{huang2020tidb} is a Raft-based HTAP database. TiSpark is a powerful analysis engine to help TiDB connect to the Hadoop ecosystem. 
The SQL engine processes process online transactions and analytical queries.
The distributed storage layer consists of a row store (TiKV) and a columnar store (TiFlash).
Two TiKV instances are deployed on two servers with a TiDB SQL engine instance.
Two TiFlash instances are deployed on the other servers with a TiSpark instance.
The TiDB~\cite{huang2020tidb} version is 5.0.0-RC, and the number of replication is two.
TiDB provides snapshot isolation or repeatable read isolation levels, and we choose repeatable read isolation in the experiments.
The MemSQL~\cite{MemSQL111} cluster consists of aggregator nodes and leaf nodes.
The aggregator nodes receive queries and distribute them to leaf nodes. 
The leaf nodes store data and execute queries issued by the aggregator nodes.
The version of MemSQL is 7.3, and the number of replication is two.
A MemSQL cluster consists of at least one master aggregator node and one leaf node.
We keep two leaf nodes, one aggregator node, one master aggregator node in 4 separate servers.
MemSQL only supports a read committed isolation level.
We remain the default configurations of the above two distributed hybrid databases.

\begin{figure}[!t]
 \includegraphics[height=3.8cm, width=9.0cm,center]{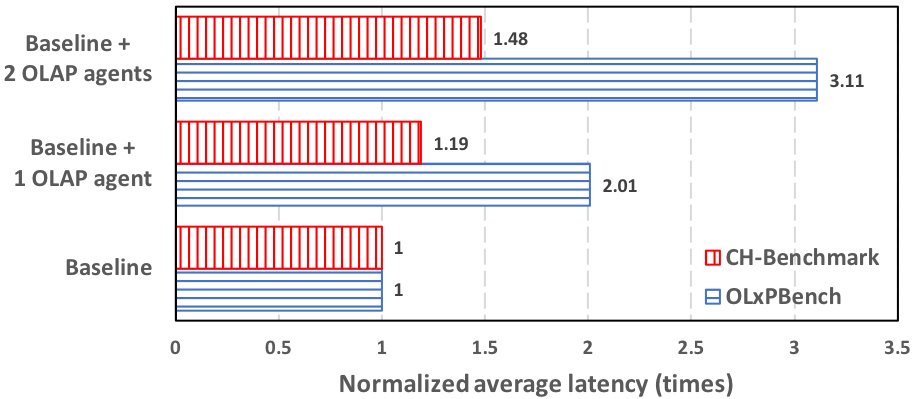}
 \caption{Comparing the schema model of OLxPBench and CH-benCHmark on TiDB cluster.}
 \label{fig: schemamodel}
 \vspace{-0.3cm}
\end{figure}
\begin{figure} 
 \includegraphics[height=3.81cm, width=8.9cm, left]{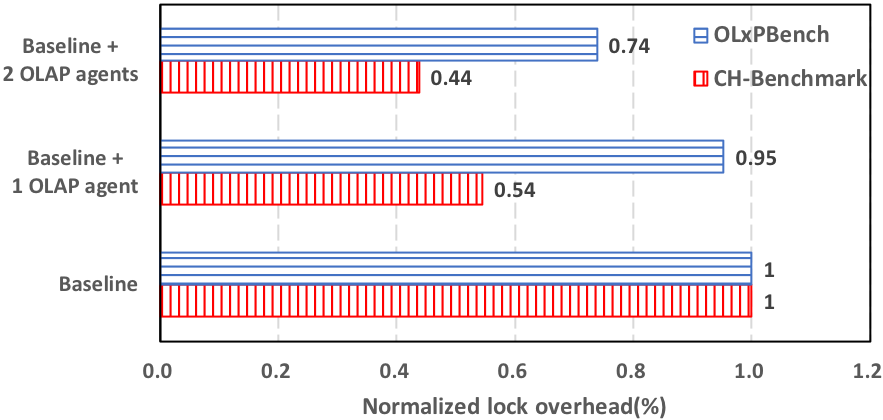}
 \caption{Comparing the lock overhead of different schema models.}
 \label{fig: lock}
\vspace{-0.5cm} 
\end{figure}
\subsubsection{Workloads Deployment}~\label{sec: 5.1.2}
The workloads in the following experiments have three sources: subenchmark, fibenchmark, and tabenchmark.
Each benchmark contributes two composites of workloads: (1) the OLTP agents and OLAP agents launch the mixtures of online transactions with analytical queries; (2) The hybrid agents send the hybrid workloads performing a real-time query in-between an online transaction to simulate the user behavior.
We do not test the performance of the cold start procedures, so there is a 60-second warm-up period before the 240-second run time.  
All the workloads are open-loop, and the request rates used in the experiment depend on the cluster's peak performance.
The requested rates vary during the experiment, and the warehouse quantity is 50.
The interference between online transactions and analytical queries increases with the increasing request rates. The transactional/analytical request rate unit is transactions per second (tps). OLxPBench reports the average latency, tail latency, and throughput.

\subsection{The evaluation of OLxPBench key design features}\label{sec: 5.3}
In this subsection, we demonstrate the design features of OLxPBench. \textbf{(1) Which schema model to adopt}, \textbf{ (2) how the real-time query impacts the performance of HTAP systems}, and \textbf{(3) Why the domain-specific benchmark should be considered}. 
The following results are the average results of the three runs.
\subsubsection{Schema Model}~\label{sec: 5.3.1}
We explore the performance difference between the semantically consistent and traditional stitched schema in the same TiDB~\cite{huang2020tidb} cluster. 
We choose the state-of-the-practice HTAP benchmark, CH-benCHmark~\cite{cole2011mixed}, as the reference.
Because CH-benCHmark~\cite{cole2011mixed} is still the most popular HTAP benchmark.
The average number of requests \textit{L} equals the long-term average arrival rate \textit{$\lambda$} multiplied by the average latency \textit{W} that a request spends in the system. The Little's Law~\cite{little2008little} is
\begin{equation}
    L=\lambda W
\end{equation}
According to Little's Law~\cite{little2008little}, the load stress in the TiDB is directly influenced by the average number of requests \textit{L} rather than the different average request arrival rates \textit{$\lambda$} of open data generator (OLxPBench) and closed data generator (CH-benCHmark). So, when the average number of requests \textit{L} in the queuing system (TiDB) is fixed, the load stress in the TiDB is fixed.
The average number of online transactions in a stable TiDB cluster over the long term is around 45.
We drop the write-heavy transactions such as \emph{NewOrder} and \emph{Payment} to reduce the possibility of load imbalance.

\textbf{Test Case 1: Varied OLAP pressures.}
The sum of the OLAP thread increases from zero to two. We put the incremental OLAP pressure on the tested systems to disturb the performance of online transactions and compare the latency of the different schema designs.
We normalized the baseline of OLxPBench and CH-benCHmark to make a fair comparison.
Little's Law does not guarantee that OLxPBench and CH-benCHmark have the same transmission rate of request, so the absolute value comparison is unfair.
Figure~\ref{fig: schemamodel} shows that the normalized average latency of online transactions in OLxPBench is more than double with the lowest OLAP pressure compared to without the OLAP pressure.
However, the normalized average latency of online transactions in CH-benCHmark increases by no more than one-fifth of the baseline under the same OLAP pressure.
Under the enormous OLAP pressure, the normalized average latency of online transactions in OLxPBench increases more than three times.
Each OLAP thread sends one OLAP query per second. 
The OLAP queries are time-consuming scan tables operations that bring a large amount of data into the buffer pool and evict an equivalent amount of older data.
Two OLAP threads can generate enormous pressure and cause a significant increase in server-side average CPU utilization.
At the same time, the normalized average latency of online transactions in CH-benCHmark increases by around 48 percent of the baseline.

TiDB~\cite{huang2020tidb} provides a row-based store, TiKV, and a column-based store,  TiFlash.
The data in TiFlash keep consistent with the data in TiKV using asynchronous log replication.
Nevertheless, the scan tables operations can occur in the row store of TiKV or the column store of TiFlash.
As the number of OLAP agents increases, the number of scan table operations increases.
We use the performance monitoring unit tool such as Perf to obtain the performance events and count the overhead of the lock.
According to the Linux Perf tool's manual, samples are performance data collected using the ' perf record' command. Lock samples indicate the number of samples collected in the lock function.
Lock overhead includes the syscall overhead of mutual exclusion (mutex) locks, fast userspace mutex (futex), and spinlock.
Lock overhead \textit{LO} equals the number of lock samples \textit{LS} divided by the total number of samples \textit{TS}.
The baseline lock overhead \textit{BLO} is the lock overhead of the online transactions without analytical query influencing.
\begin{figure}[!t]     
 \includegraphics[height=3.80cm, width=9cm, center]{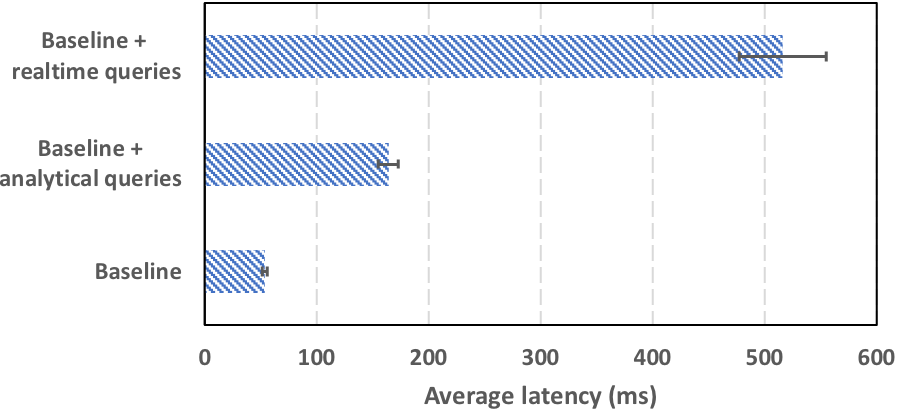}
 \caption{Comparing the analytical queries to the real-time queries of subenchmark on the TiDB cluster.}
 \label{fig: query}
 \vspace{-0.5cm}
\end{figure}
The normalized lock overhead \textit{NLO} is the lock overhead \textit{LO} divided by the baseline lock overhead \textit{BLO}.
\begin{equation}
    NLO = \frac{LS}{TS*BLO} \times 100\%
\end{equation}
When the analytical agent increases, the throughputs of online transactions will be influenced.
So the normalized lock overhead decreases with the analytical agent increases in Figure~\ref{fig: lock}.  
The difference in performance isolation measured by OLxPBench is far more significant than CH-benCHmark. Better performance isolation indicates that the execution of OLTP with OLAP workloads affects the other one’s performance much lighter. 
Figure~\ref{fig: lock} reports that the lock overhead gap between semantically consistent schema and stitched schema is 1.76x using one OLAP thread and 1.68x using two OLAP threads.
It indicates that the shared data between OLTP and OLAP on a semantically consistent schema is more significant than stitched ones. 
The low competition in CH-benCHmark between OLTP and OLAP workloads will propagate a false image that the HTAP system can guarantee isolated performance.

\textbf{Implication 1}
\emph{Experiments show that semantically consistent schema reveals inherent competition between OLTP and OLAP than stitched schema.}
\begin{figure}[!t]
 \includegraphics[height=3.8cm, width=8cm, left]{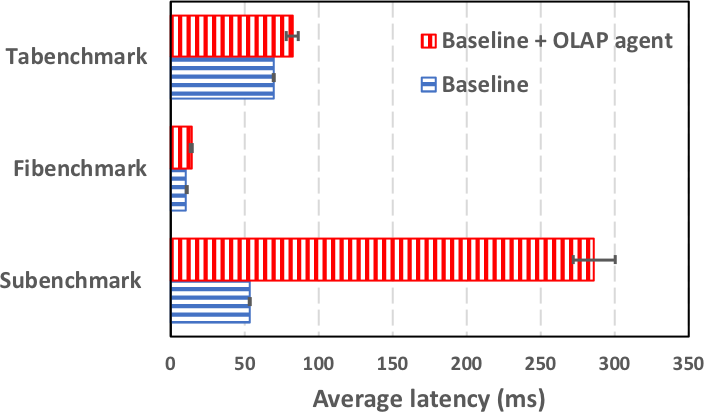}
 \caption{Comparing the generic benchmark to the domain-specific benchmarks on TiDB cluster.}
 \label{fig: benchmarks}
 \vspace{-0.5cm}
\end{figure}

\subsubsection{Real-time Query}~\label{sec: 5.3.2}
We now compare the two main queries common in real-world scenarios: \emph{analytical queries} and \emph{real-time queries}.
First, the analytical queries keep the essential characteristics such as the complexity of operations and perform multidimensional customer analysis.
Second, the real-time queries in OLxPBench are extracted from the existing production applications.  
The real-time queries are used in the production applications to perform real-time user behavior simulations.
However, none state-of-the-art~\cite{coelho2017htapbench} and state-of-the-practice~\cite{cole2011mixed} HTAP benchmarks provide the workloads that include real-time queries imitating user behavior.
We compare the two different intentional queries on the TiDB cluster.

\textbf{Test Case 2: Queries comparison.} 
We run the subenchmark using the semantically consistent schema at 30 online transactions per second as the baseline.
Then we inject analytical queries at 1 query per second into the baseline as experimental group one.
Meanwhile, we send the hybrid transaction performing a real-time query in-between an online transaction at 30 requests per second as the experimental group two.
Figure~\ref{fig: query} shows that the analytical queries increase the baseline latency by around three times.
The real-time queries in hybrid transactions increase the baseline latency by more than nine times.
The hybrid transaction contains both OLTP statements and OLAP statements, but the SQL engine can only choose a row-based store or column-based store to handle the hybrid transaction.
However, the analytical queries and the online transactions can be handled separately by the column-based TiFlash and the row-based TiKV.
Therefore, the impact of real-time query simulating the user behavior is more significant than the impact of analytical queries on the performance of online transactions.
Besides, the standard deviation of the average baseline latency is 2.21.
With the analytical queries interference, the standard deviation of average baseline latency increases from 2.21 to 9.16.
Under the real-time queries interference, the standard deviation of average baseline latency increases from 2.21 to 38.91.
It indicates that interference of real-time queries to online transactions is greater than that of analytical queries. 
\begin{figure*} \centering   
\subfigure[Throughput of OLTP.] {
 \label{fig.suoltp}     
\includegraphics[height=3.3cm, width=5.5cm]{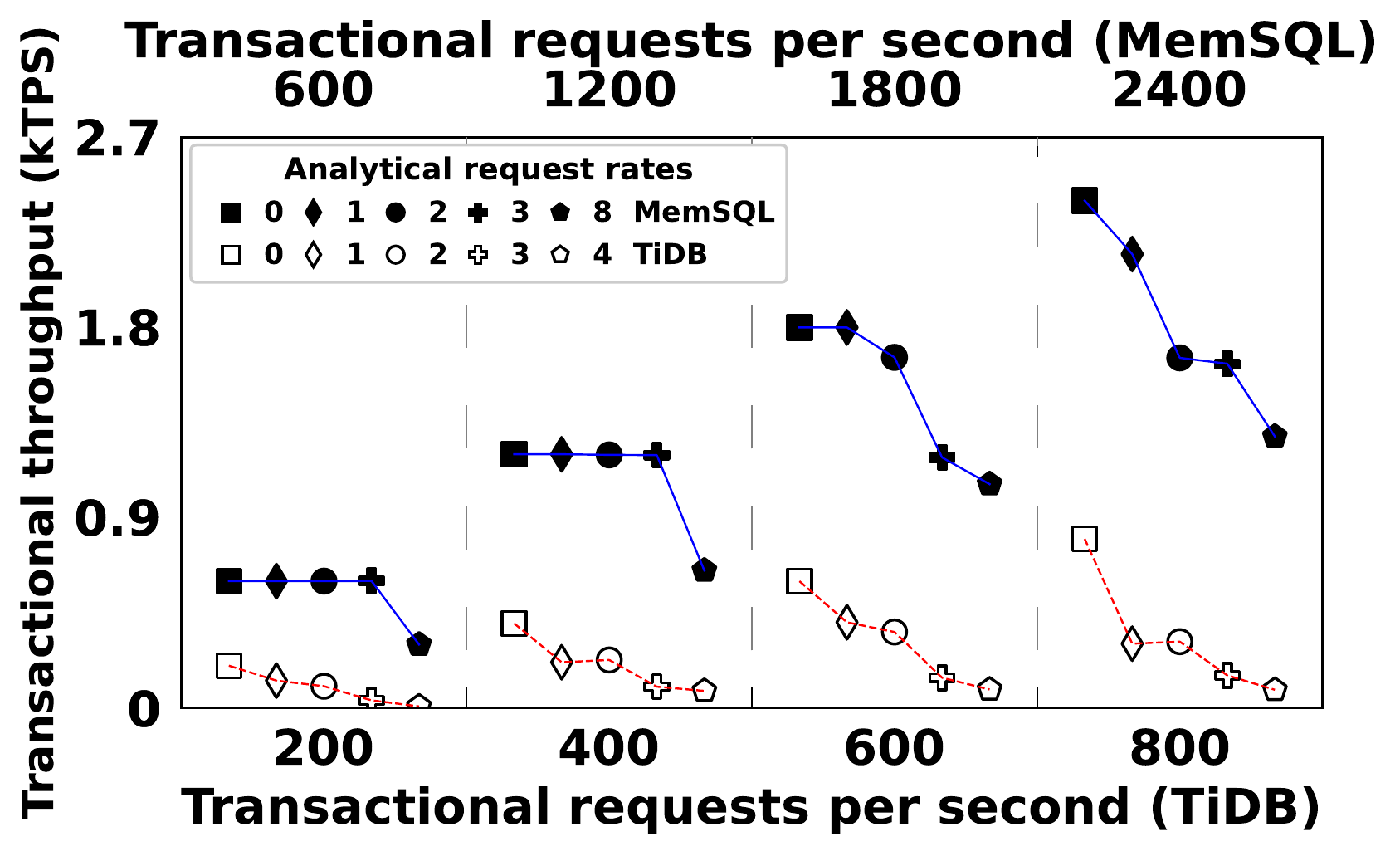}  
}     
\subfigure[Throughput of OLAP.] {  
\label{fig.suolap}     
\includegraphics[height=3.3cm, width=5.4cm]{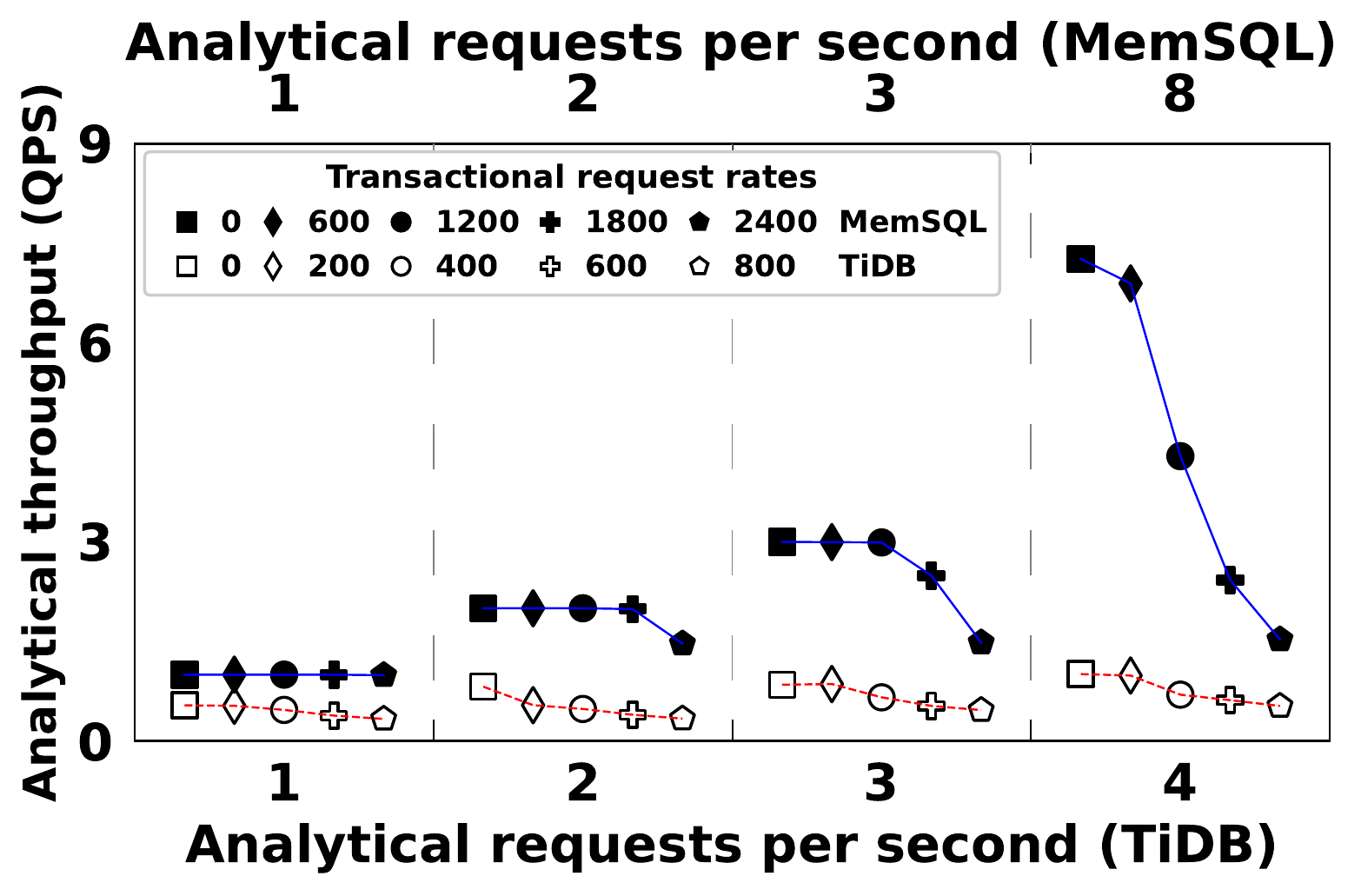}    
} 
\subfigure[Throughput of OLxP.] { 
\label{fig.suolxp}     
\includegraphics[height=3.3cm, width=5.4cm]{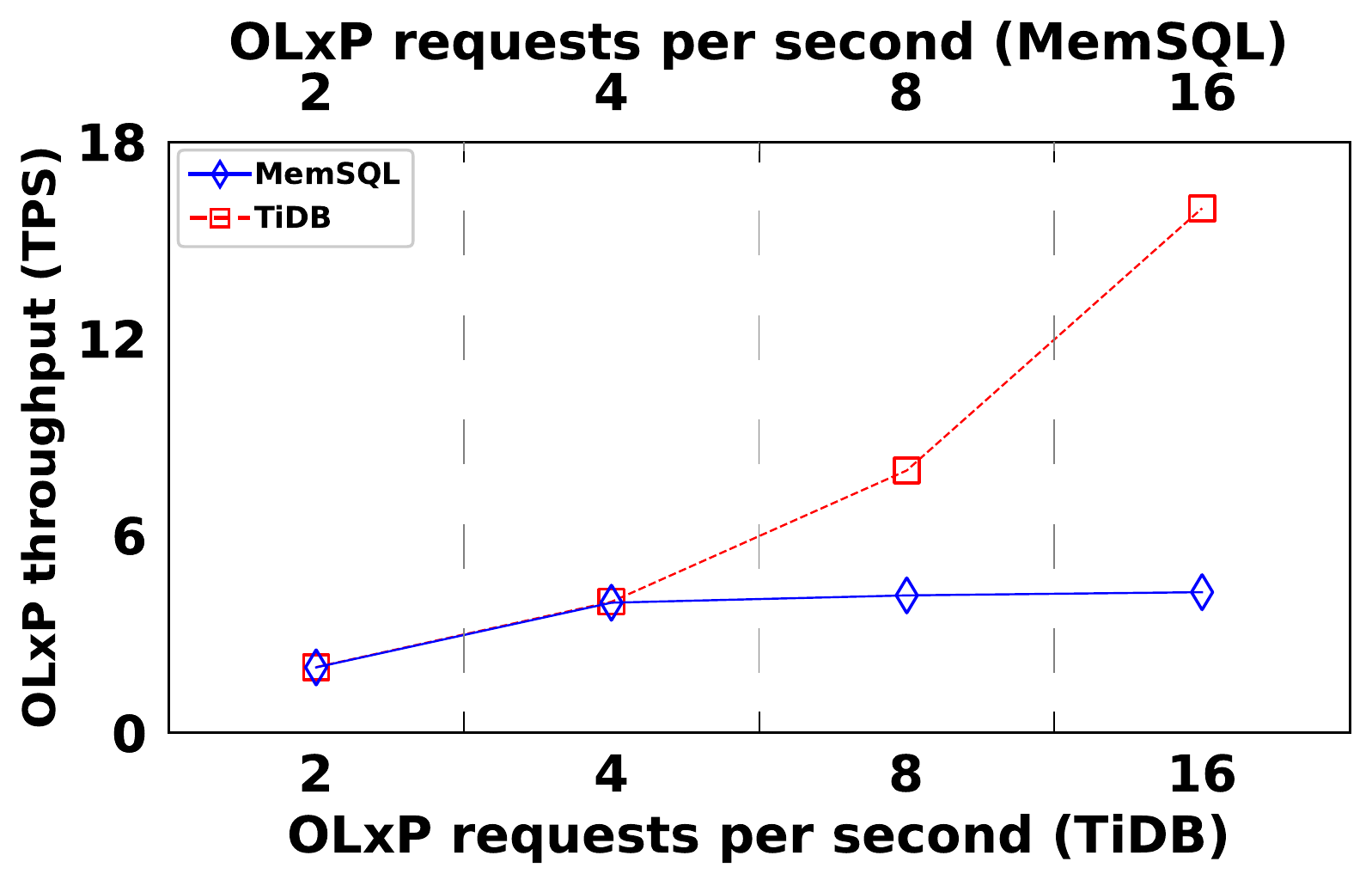}   
}  
\caption{ OLTP, OLAP and OLxP performance of subenchmark. }     
\label{fig.su}  
\vspace{-0.3cm}
\end{figure*}
So it is necessary to include the real-time queries extracted from the production environment in the HTAP benchmark for helping users choose the appropriate HTAP system to handle real-time queries.

\textbf{Implication 2}
\emph{It is necessary to include the real-time queries
in the HTAP benchmark for testing whether the HTAP system can handle real-time queries from users.}

\subsubsection{Domain-specific Benchmark}~\label{sec: 5.3.3}
In this paper, we classify the benchmarks into two categories: \emph{generic benchmark} and \emph{domain-specific benchmark}.
The subenchmark is inspired by TPC-C~\cite{TPCC111}, which is not bound to a specific scenario, and the community considers a general benchmark. The fibenchmark and the tabenchmark model a banking scenario and a telecom scenario. 
Hence, they are domain-specific benchmarks.

\textbf{Test Case 3: Domain-specific benchmark.}
We run the subenchmark, the fibenchmark, and the tabenchmark at 80 online transactions per second as the baseline.
Then we send the analytical queries at 1 query per second with the baseline.
Figure~\ref{fig: benchmarks} shows that the baseline of the above three benchmarks is 53.47ms, 10.25ms, and 69.53ms.
Moreover, the standard deviations of the baseline of the above three benchmarks are 0.23, 0.05, and 0.47.
The online transactions of fibenchmark perform read-heavy and simple update operations, so its baseline latency is the smallest one.
Slow queries took longer than one second in tabenchmark's online transactions. So the baseline of the tabenchmark is the biggest one.
We will analyze the reason for slow queries in Section~\ref{sec: 6.3}.
Only 8\% of online transactions in subenchmark do not modify the table, and the tables in subenchmark contain complex relations. 
So its baseline average latency is the median.

Under the OLAP pressure, the OLTP latency of subenchmark increases by more than five times, and the OLTP latency of fibenchmark increases by less than forty percent.
And the OLTP latency of tabenchmark increases by less than twenty percent.

\begin{figure*} \centering   
\subfigure[Throughput of OLTP.] {
 \label{fig.fioltp}     
\includegraphics[height=3.3cm, width=5.5cm]{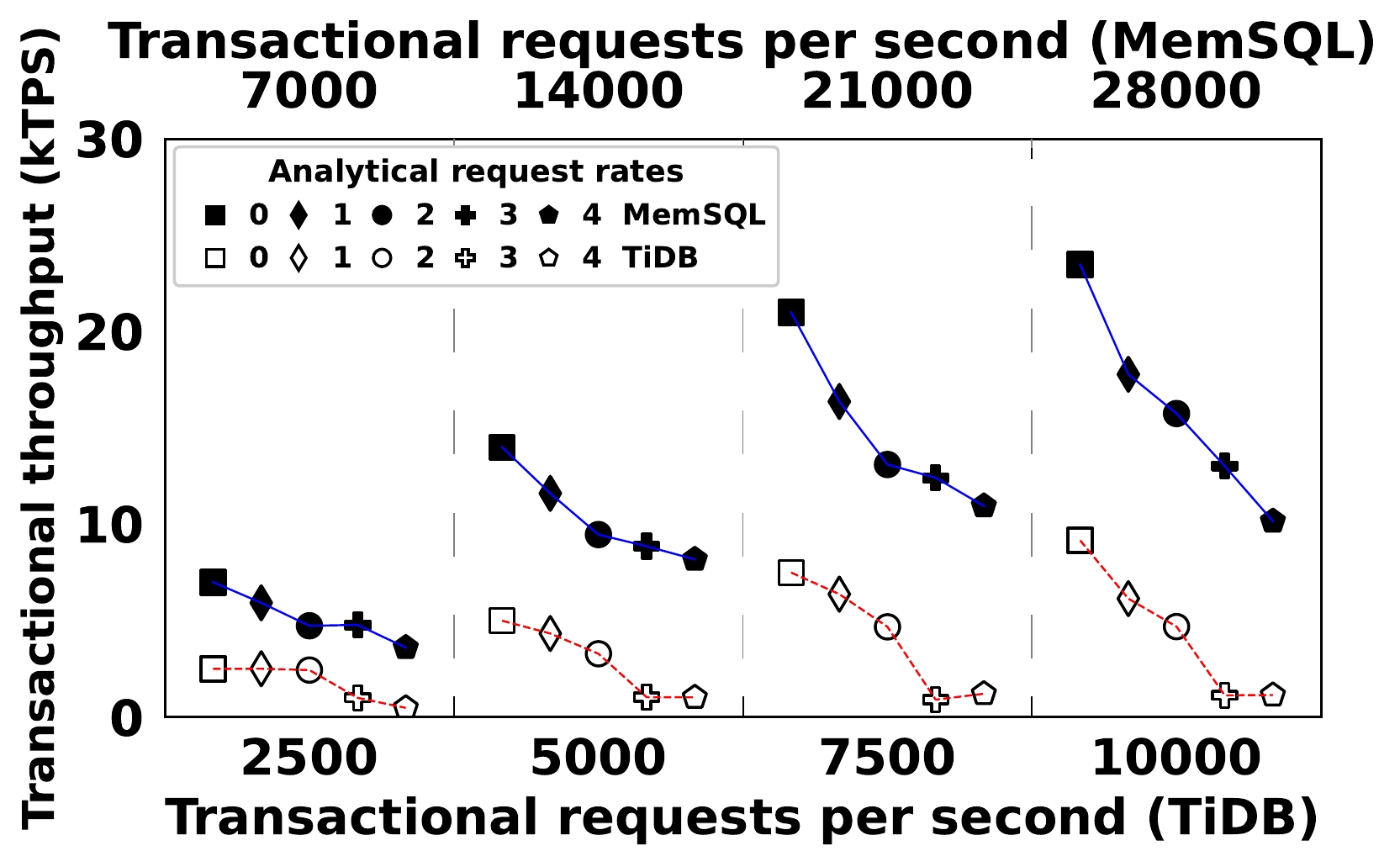}  
}     
\subfigure[Throughput of OLAP.] { 
\label{fig.fiolap}     
\includegraphics[height=3.3cm, width=5.6cm]{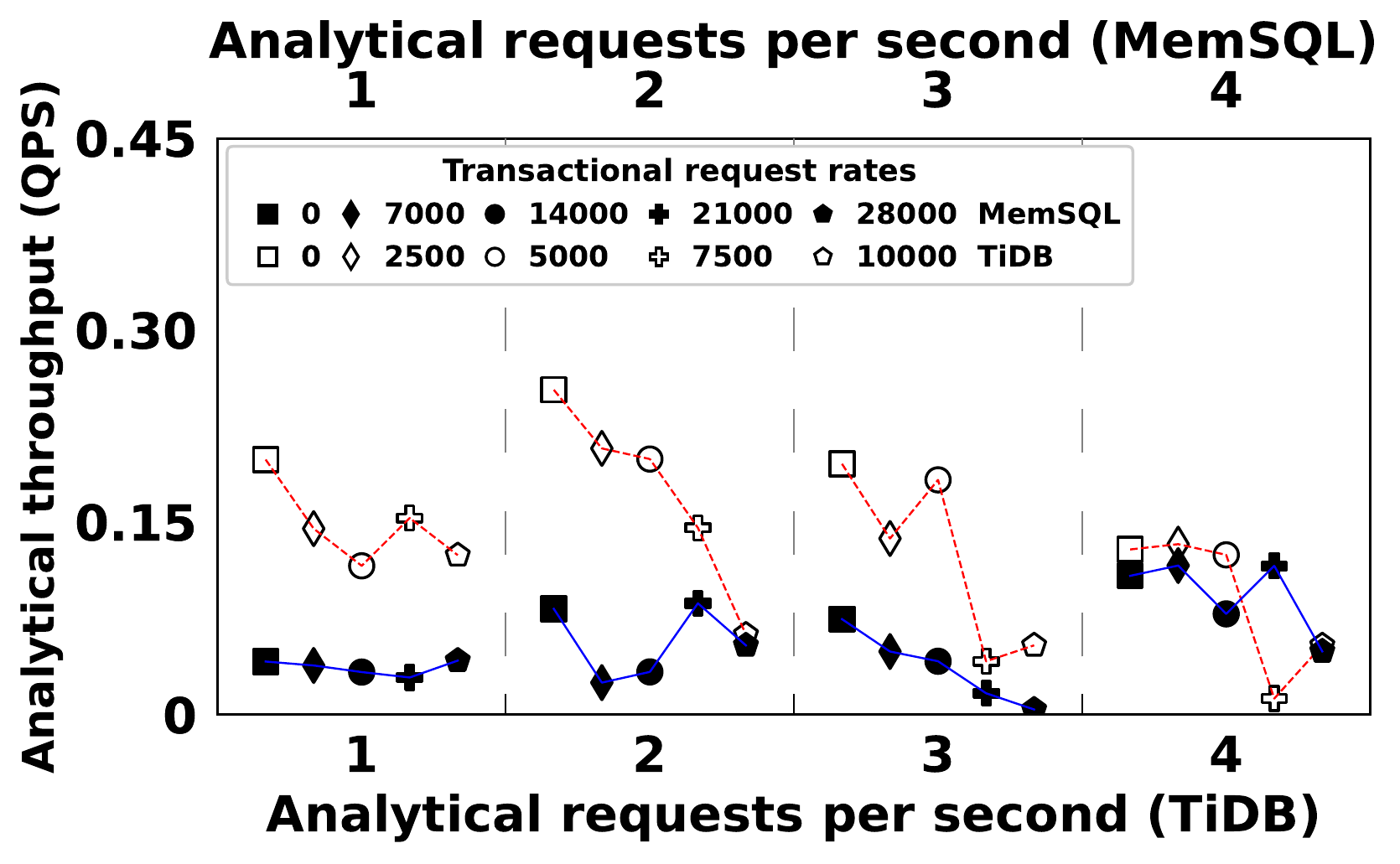}    
} 
\subfigure[Throughput of OLxP.] { 
\label{fig.fiolxp}     
\includegraphics[height=3.3cm, width=5.4cm]{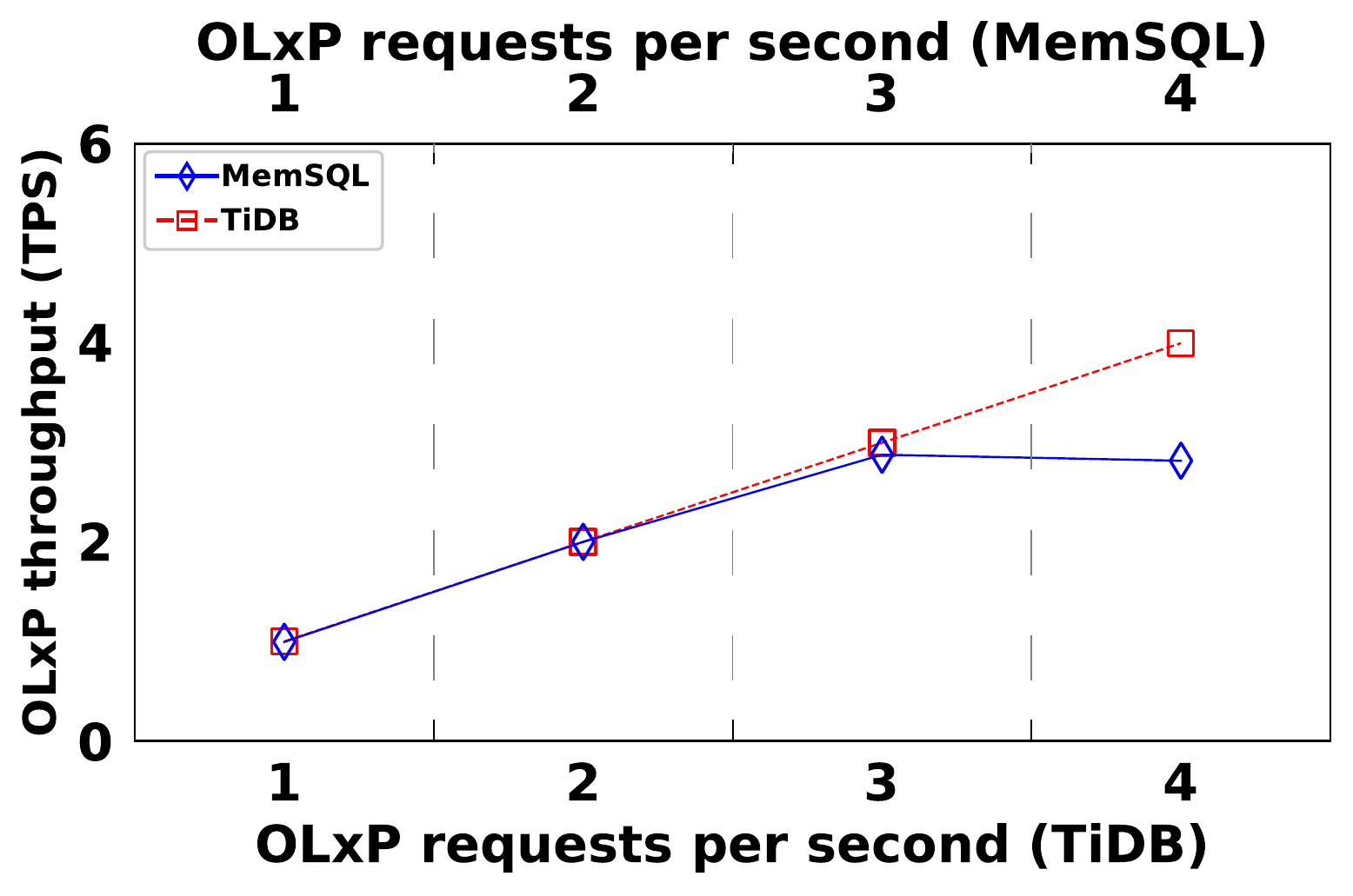}   
}  
\caption{ OLTP, OLAP and OLxP performance of fibenchmark. }     
\label{fig.fi}  
\vspace{-0.3cm}
\end{figure*}

Meanwhile, the standard deviations of the above three benchmarks increase to 14.10, 0.58, and 4.05 under the analytical queries interference.
The complex analytical queries in subenchmark generate many table scan operations, which increase the waiting time of online transactions.
The read-heavy online transactions of fibenchmark are mostly negligible by OLAP agents.
Therefore, online transactions of subenchmark are most affected by OLAP pressure, followed by fibenchmark' online transactions, and tabenchmark' online transactions are the least affected.

\textbf{Implication 3}
\emph{The domain-specific benchmarks help users identify system bottlenecks in their specific scenarios.
Besides, it also helps system designers point in the direction of system optimization.  }

\section{Evaluation of The Mainstream Distributed HTAP DBMSs}\label{sec: 6}

In addition to the feature evaluation of OLxPBench, we also thoroughly test end-to-end performance for mainstream distributed HTAP DBMSs.
We will provide detailed evaluation data in the following subsections, including peak performance.
The peak performance refers to the saturation value that a single workload can reach in the test cluster. 
Furthermore, we will describe and deeply analyze the mutual interference between OLTP and OLAP~\cite{9458644} using the control variate method.
The transactional/analytical request rates are divided into four numerically increasing groups with the same interval based on peak throughput.
The transactional/analytical request rates in each group are the same, and the analytical/transactional request rates increase from zero to peak to explore the influence of the analytical/transactional agents on the transactional/analytical agents.
Besides, CH-benCHmark~\cite{cole2011mixed} uses the stitch schema while OLxPBench uses a semantically consistent one. In addition, OLxPBench uses hybrid workloads.
The difference in performance isolation measured by OLxPBench is far more significant than CH-benCHmark. Better performance isolation indicates the execution of OLTP with OLAP workloads affects the other one’s performance much lighter. Moreover, we find that the lock overhead gap between the OLxPBench and CH-benCHmark is 1.76x under the same OLAP pressure in TiDB. The low competition in CH-benCHmark between OLTP and OLAP workloads will propagate a false image that the HTAP system can guarantee isolated performance.
So, we do not report the experimental results of CH-benCHmark.

\subsection{Subenchmark evaluation}\label{sec: 6.1}
\subsubsection{Peak performance\label{sec: 6.1.1}}
Figure~\ref{fig.suoltp} shows that the transactional throughput increases with the incremental transactional request rates.
In the MemSQL cluster, the throughput reaches the top when the transactional request rates are 2400 tps.
The average latency of transactions is 29.7 milliseconds without OLAP agent inferences.
And the 95th percentile latency of transactions is 78.53 milliseconds.
In the TiDB cluster, the maximum transactional throughput is 800 tps.
Figure~\ref{fig.suoltp} illustrates that the throttled transactional throughput of subenchmark in the TiDB cluster is one-third that of the MemSQL cluster. 
The above result is the data processing of MemSQL in memory rather than in solid-state disk.
Figure~\ref{fig.suolap} shows that the maximum analytical throughput is around eight tps in the MemSQL cluster.
Moreover, the analytical throughput reaches the top when the analytical request rates are four tps in the TiDB cluster.
Under the same analytical request rates, the average latency of OLAP increases with the incremental transactional request rates.

The performance of OLxP in the different hybrid request rates is shown in Figure~\ref{fig.suolxp}.
The OLxP workloads include hybrid transactions, which perform a real-time query in-between an online transaction to simulate the user behavior.
The real-time query is a time-consuming aggregate operation.
So, the transactional statements behind the real-time query must wait for the real-time query execution because of the atomicity property of the transaction. So, the maximum throughput of OLxP is 4.28tps and 15.98tps in Figure~\ref{fig.suolxp}.
In the MemSQL cluster, the maximum average hybrid latency is 133.44 seconds.
The gap between the maximum and minimum average delays is 223 times.
And the 95th percentile latency of hybrid workload is 209.50 seconds.
MemSQL adopts the vertical partitioning technology, which results in many join operations generated by relationship query statements in hybrid transactions and increases the waiting time of hybrid transactions.
In the TiDB cluster, the throttled OLxP throughput is 16 tps, and the maximum average hybrid latency is 397 milliseconds.
The maximum average delay is  1.47 times the minimum average delay.
And the 95th percentile latency of hybrid workload is 905.36 milliseconds.
The above results indicate that TiDB's separated storage engine can handle the OLxP workloads compared to MemSQL's single storage engine.

\subsubsection{Performance interference between OLTP agents and OLAP agents\label{sec: 6.1.2}}
The performance impact of analytical agents on transactional agents is shown in Figure~\ref{fig.suoltp}.
When the transactional request rates are controlled, the average latency of the transactional agents increases by up to 17.4 times compared with the absence of the analytical agents in the MemSQL cluster.
And the gap of the 95th percentile latency is 33.7x.
The performance impact of the online transactions on the analytical queries is shown in Figure~\ref{fig.suolap}.
When the analytical request rates are controlled, the average latency of the analytical agents increases by up to 2.2 times compared with the absence of the transactional agents.
And the gap of the 95th percentile latency is 2.0x.
It indicates that violent interference exists between transactional agents and analytical agents.
The expensive analytical queries compete for the resource with the online transactions in the single storage and increase the latency of online transactions. 

In the TiDB cluster, under the same analytical request rates, the analytical throughput decreases to the baseline of 59\% as the transactional request rate increases.
Furthermore, when the transactional request rates are 800 tps, the transactional throughput plummets as the analytical request rates increase, up to 89\%.
The transactional agents significantly affect the execution of analytical agents.
The higher the request rates, the more table scan operations.
Time-consuming table scan operations increase the waiting time of requests and reduce requests throughput.

\subsection{Fibenchmark evaluation}\label{sec: 6.2}
\subsubsection{Peak performance\label{sec: 6.2.1}} 
As shown in Figure~\ref{fig.fioltp}, the peak transactional throughput is around 23476 tps in the MemSQL cluster.
The maximum transactional throughput is 9165 tps in the TiDB cluster.
The read-only transaction ratio of fibenchmark is higher than that of subenchmark, so the peak transactional throughput of fibenchmark is higher than that of subenchmark.
A large number of queries are blocked until the previous complex queries are completed.
So the peak analytical throughput of MemSQL is around 0.12 tps in Figure~\ref{fig.fiolap}.
And the maximum analytical throughput of TiDB is 0.25 tps. 
There are a lot of scan table operations in the workloads of fibenchmark, and scanning row-format tables in TiKV~\cite{huang2020tidb} is stochastic and expensive, explained in Section~\ref{sec: 5.3.1}.
Figure~\ref{fig.fiolxp} shows that the hybrid throughput increases as the hybrid request rates increase when the hybrid request rates are no more than four tps.
In the MemSQL cluster, the peak hybrid throughput is 2.9 tps.
In the TiDB cluster, the peak hybrid throughput is 4 tps.
The average hybrid latency increases at most 17.2\% as the hybrid request rates increase.
And the 95th percentile latency increases up to 36.4\% as the hybrid request rates increase.
The hybrid latency increase as the hybrid request rates increase without bound.
The increasing average and 95th percentile latency of hybrid transactions result from more waiting time with the higher hybrid request rates.

\begin{figure*} \centering   
\subfigure[Throughput of OLTP.] {
 \label{fig.taoltp}     
\includegraphics[height=3.4cm, width=5.6cm]{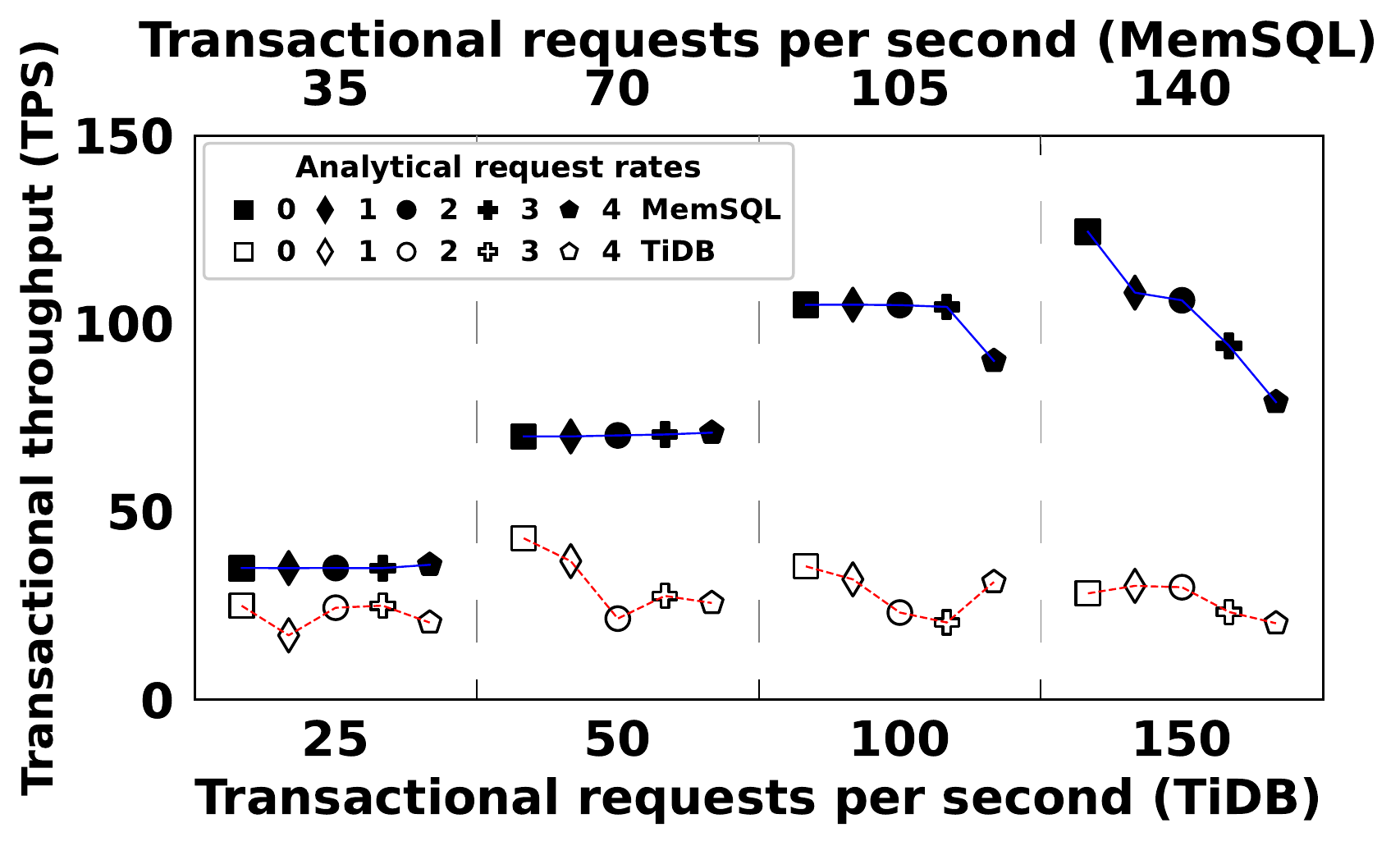}  
}     
\subfigure[Throughput of OLAP.] { 
\label{fig.taolap}     
\includegraphics[height=3.4cm, width=5.5cm]{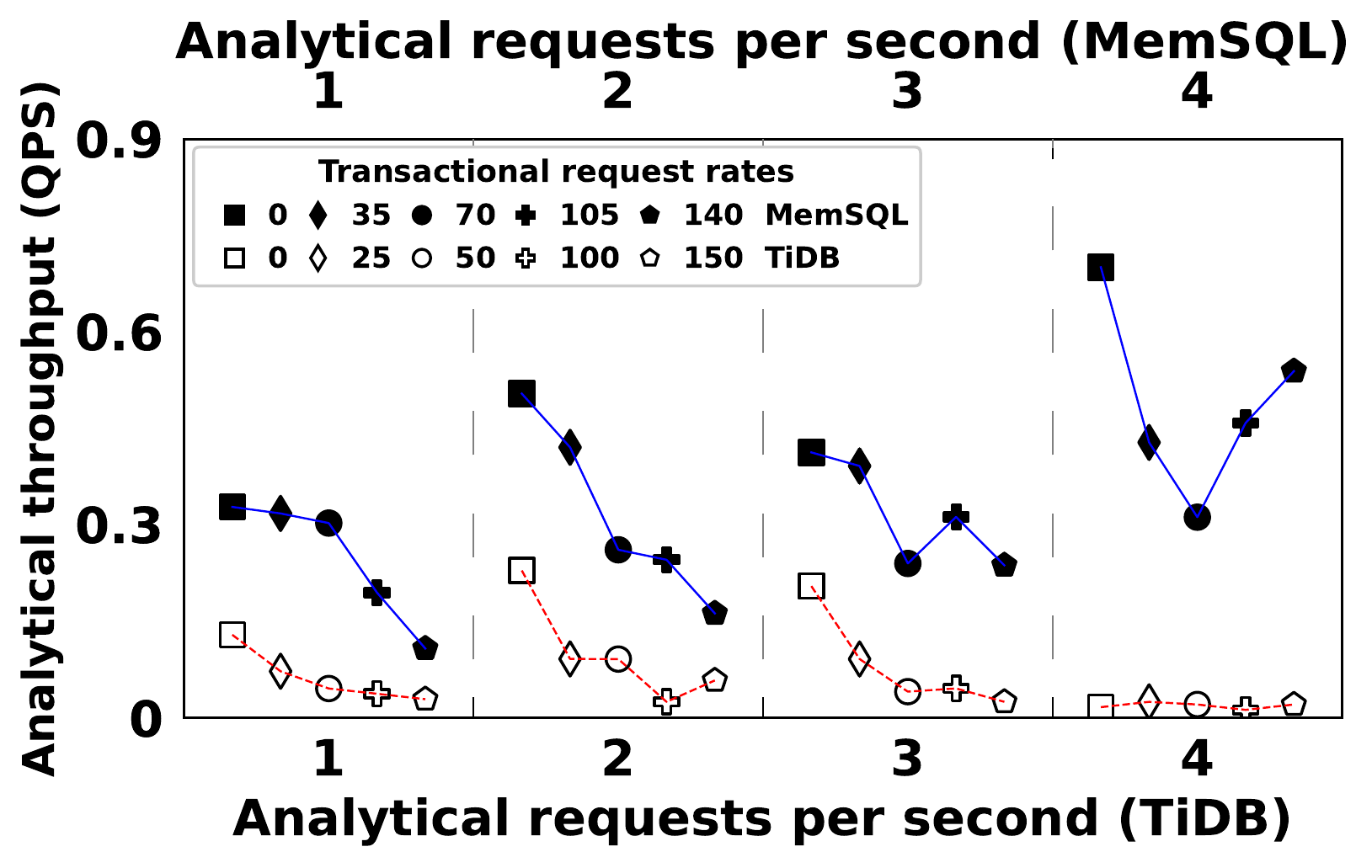}    
}
\subfigure[Throughput of OLxP.] { 
\label{fig.taolxp}     
\includegraphics[height=3.4cm, width=5.5cm]{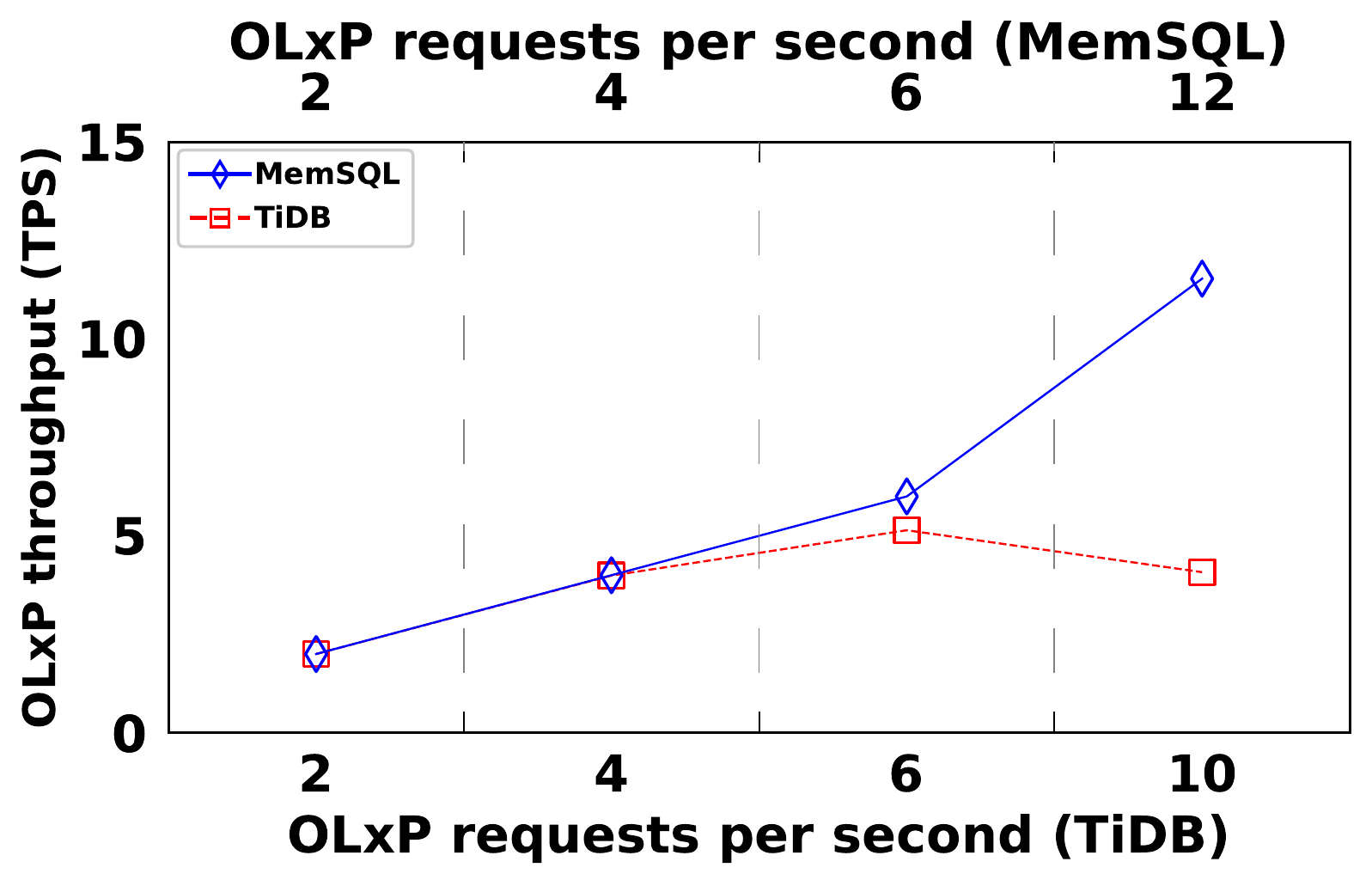}   
}  
\caption{ OLTP, OLAP and OLxP performance of tabenchmark. }     
\label{fig.ta}  
\vspace{-0.3cm}
\end{figure*}

\subsubsection{Performance interference between OLTP agents and OLAP agents\label{sec: 6.2.2}} 
The performance impact of analytical agents on transactional agents is shown in Figure~\ref{fig.fioltp}.
And the performance impact of transactional agents on analytical agents is shown in Figure~\ref{fig.fiolap}.
In the MemSQL cluster, the transactional throughput decreases with the analytical request rates increasing under the same transactional request rates.
And the average latency of the transactional request rates increases as the analytical request rates increase.
The analytical throughput decreases with the incremental transactional request rates when the analytical request rates are less than three tps, and the transactional request rates are less than 7000 tps.
The analytical throughput fluctuates wildly when the analytical request rates exceed the processing capacity of the MemSQL.
Meanwhile, the long-term running analytical queries increase the waiting time of online transactions.

In the TiDB cluster, under the same transactional request rates, the transactional throughput decreases as the analytical request rates increase when the analytical request rates are no more than 3 tps.
Under the same analytical request rates, the analytical throughput decreases as the transactional request rates increase when the analytical request rates are no more than 2 tps, and the transactional request rates are no more than 5000 tps.
And the average analytical latency increases as the transactional request rates increase when the transactional request rates are no more than 2500 tps.

\subsection{Tabenchmark evaluation}\label{sec: 6.3te}
\subsubsection{Peak performance\label{sec: 6.3.2}}
Figure~\ref{fig.taoltp} shows that the maximum transactional throughput is 124 tps in the MemSQL cluster.
The transactional throughput increases as the transactional request rates increase when the transactional request rate is no more than 140 tps.
The maximum transactional throughput is 43 tps in the TiDB cluster.
The transactional throughput increases as the transactional request rates increase when the transactional request rates are no more than 50 tps.
Tabenchmark has the highest percentage of read-only transactions among the three benchmarks.
However, there is a slow query in the \textit{DeleteCallForwarding} transaction that took longer than one second, so tabenchmark has the lowest transactional throughput of the three benchmarks.
The SQL statement is \textit{"explain SELECT s\_id FROM SUBSCRIBER WHERE sub\_nbr = ?"}.
The s\_id and sub\_nbr are the composite keys of the SUBSCRIBER table.
The full table scan in memory is time-consuming when the slow query is executed in MemSQL.
Even worse, when a slow query is executed in the storage engine of the TiDB, the index full scan will perform a random read on the solid-state disk.
Therefore, the maximum transactional throughput of MemSQL is higher than the maximum transaction throughput of TiDB.
Figure~\ref{fig.taolap} shows that the maximum analytical throughput is 0.7 tps in MemSQL cluster.
The analytical throughput increases as the analytical request rates increase when the analytical request rates are no more than two tps.
And the maximum analytical throughput is 0.23 tps in the TiDB cluster.
The analytical throughput increases as the analytical request rates increase when the analytical request rates are no more than two tps.

Figure~\ref{fig.taolxp} shows that the MemSQL cluster is saturated when the hybrid request rate increases to 12 tps.
Figure~\ref{fig.taolxp} shows that the maximum hybrid throughput is around five tps in the TiDB cluster.
And the average latency increases with hybrid request rates increases without bound.
The gap between 95th percentile latency and average latency is up to 2.2x. 

\subsubsection{Performance interference between OLTP agents and OLAP agents\label{sec: 6.3.3}}
Figure~\ref{fig.taoltp} shows that the performance impact of analytical agents on transactional agents.
The performance impact of transactional agents on analytical agents is shown in Figure~\ref{fig.taolap}.
In the MemSQL cluster, OLxPBench executes the precisely transactional request rates control when the transactional request rates are no more than 105 tps, and the analytical request rates are no more than three tps.
Under the same transactional request rates, the average transactional latency increases more than 34.4 times.
And the 95th percentile latency increases by 12.8x.
The analytical throughput decreases when the analytical request rates are identical as the transactional request rates rise, which are no more than 50 tps.

The transactional throughput decreases to 49.8\% with the analytical agents' inference in the TiDB cluster.
It indicates the analytical agents significantly increase the online transaction waiting time.
The performance impact of transactional agents on the analytical agents is shown in Figure~\ref{fig.taolap}.
The analytical throughput decreases up to 89\% under the transactional agents' inference.
The average latency increases 30.8\% under the transactional agents' inference.
And the 95th percentile latency increases 12.2\% under the transactional agents' inference.
It indicates that the slow queries in the online transaction block the analytical agents' execution.

\begin{figure*} \centering   
\subfigure[OLTP latency.] {
 \label{fig.scaleoltp}     
\includegraphics[height=3.3cm, width=5.72cm]{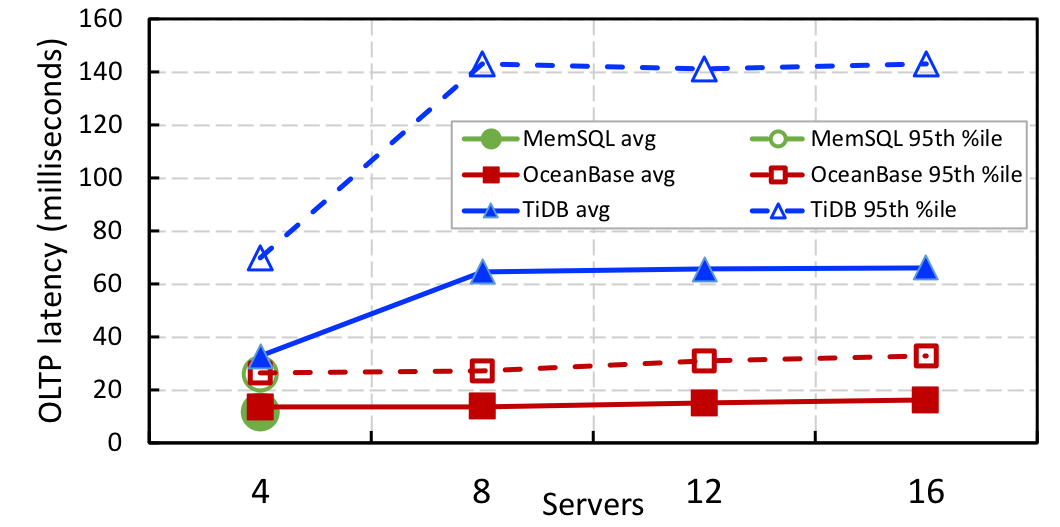}  
}     
\subfigure[OLTP latency with OLAP interference.] { 
\label{fig.scaleolap}     
\includegraphics[height=3.3cm, width=5.72cm]{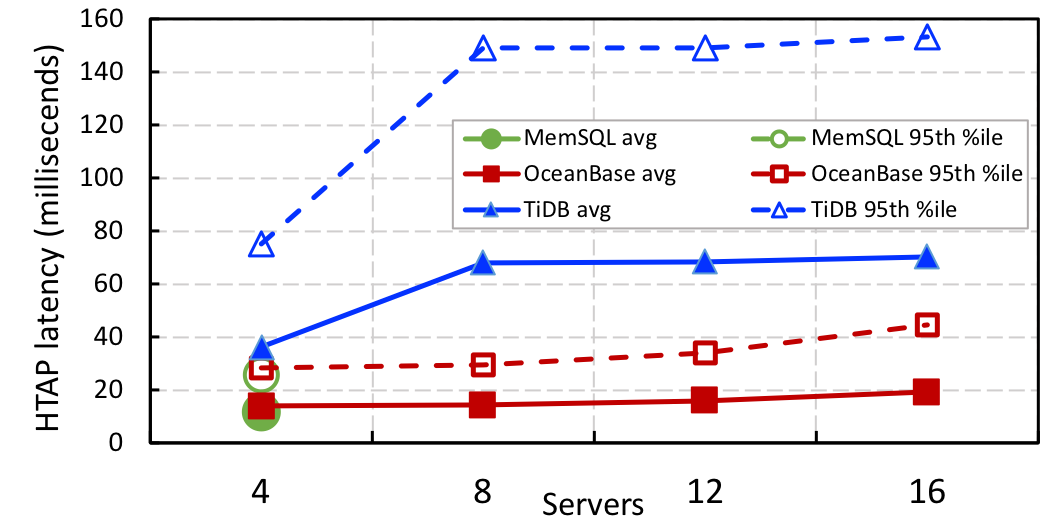}    
}
\subfigure[OLxP latency.] { 
\label{fig.scaleolxp}     
\includegraphics[height=3.3cm, width=5.72cm]{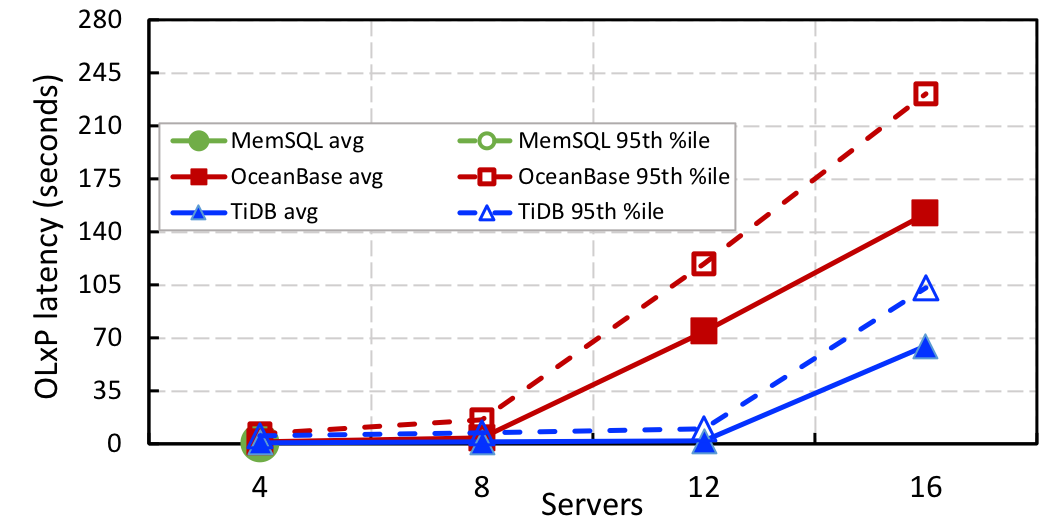}   
}  
\caption{ OLTP, HTAP and OLxP latency as cluster size increases. }     
\label{fig.scale16}  
\vspace{-0.35cm}
\end{figure*}

\subsection{The main findings of differences between MemSQL and TiDB}\label{sec:find}
First, the enormous transactional performance gap between MemSQL and TiDB results from the different storage mediums for data processing, i.e., memory for MemSQL and solid-state disk for TiDB. The peak transactional throughput gap between MemSQL and TiDB is 3.0x, 2.6x, and 2.9x using subenchmark, fibenchmark, and tabenchmark.
Second, compared to the single storage engine of MemSQL, the separated storage engines (row-based and column-based) of TiDB handle the hybrid workload better, performing real-time queries in-between an online transaction. The peak hybrid workload throughput gap between TiDB and MemSQL is 3.7x and 1.4x using subenchmark and fibenchmark.
Third, both MemSQL and TiDB handle the query using the composite keys awkwardly. MemSQL uses time-consuming full table scans in memory, while TiDB uses index full scans that perform a random read on the solid-state disk. The maximum hybrid workload throughput of MemSQL is 2.2x than that of TiDB.

\subsection{Scalability}\label{sec: 6.3}
We choose the mainstream HTAP DBMSs -- TiDB~\cite{huang2020tidb}, MemSQL~\cite{MemSQL111}, and OceanBase~\cite{OceanBase111} for scale-out experiments.
We test the scaling capability of TiDB and OceanBase by varying the cluster sizes from 4 to 16 nodes~\footnote{For too high cost for commercial software, we test MemSQL on four servers.}.
Meanwhile, the data size and target request rates rise in proportion to the increasing cluster size.
The following results are the average results of the five runs.
TiDB decouples the computational engine layer and storage layer.
Due to complex execution plans and compute-intensive queries, we set the ratio of storage instances(SI) to computational instances(CI) at 2:1 in the TiDB cluster.
Storage instances are deployed on all servers in the cluster, and computational instances are deployed on half of the servers in the cluster.
Oceanbase is shared-nothing architecture, and each OceanBase server (OBServer) is the same.
The number of OBServers is the cluster size.

The average latency and 95th percentile latency for workloads in subenchmark are shown in Figure~\ref{fig.scale16}.
First, OLxPBench clients run on a separate eight vCPU machine and can spawn up to 300 threads to generate target request rates.
OLxPBench clients can be deployed on separate client servers, so client servers are not a bottleneck.  
The more OLxPBench clients are deployed, the more requests are generated.
Second, OceanBase and TiDB cannot scale-out well when dealing with the OLTP workloads, OLxP workloads, and the mixtures of OLTP and OLAP workloads.
In the OceanBase cluster, the average latency and 95th percentile latency of OLTP workloads increase by 20\% and 24\% as the cluster size increase from 4 to 16 nodes.
In the TiDB cluster, the average latency and 95th percentile latency of OLTP workloads increase more than 1x as the cluster size increase from 4 to 16 nodes.
Significantly, the latency of OLxP workloads increases sharply as the cluster size increase from 4 to 16 nodes.
It is challenging for the above HTAP DBMSs to deal with the OLxP workloads. 
Third, compared with OceanBase, TiDB provides better performance isolation as the cluster size increases.
Under the same OLAP pressure, the average latency of OLTP workloads increases by 6\% and 18\% in the TiDB cluster and OceanBase cluster.
Meanwhile, TiDB is better than OceanBase at dealing with OLxP workloads.
The performance isolation benefits from the decoupled storage layer consisting of a row store (TiKV) and a columnar store (TiFlash). 

\section{Conclusions}
This paper quantitatively discloses that the previous HTAP benchmarks provide misleading information in evaluating, designing, and implementing HTAP systems.  
We design and implement an extensible HTAP benchmarking framework named OLxPBench.  OLxPBench proposes the abstraction of a hybrid transaction to model the widely-observed behavior pattern -- making a quick decision while consulting real-time analysis; a semantically consistent schema to express the relationships between OLTP and OLAP schemas;  the combination of domain-specific and general benchmarks to characterize diverse application scenarios with varying resource demands.  
Extensive experiments demonstrate its merit and pinpoint the bottlenecks of the mainstream distributed HTAP DBMSs.

\bibliographystyle{IEEEtran}

\end{document}